\documentclass[12pt]{article}

\usepackage{amsmath,amsfonts,amsthm,amssymb}
\usepackage{xcolor}

\swapnumbers 

\newtheorem{lemma}{Lemma}[section]
\newtheorem{corollary}[lemma]{Corollary}
\newtheorem{theorem}[lemma]{Theorem}
\newtheorem{prop}[lemma]{Proposition}

\theoremstyle{remark}
\newtheorem{example}[lemma]{Example}

\theoremstyle{definition} 
\newtheorem{definition}[lemma]{Definition}

\newtheorem{assumption}[lemma]{Standing Hypothesis}

\newcommand{\face}{\operatorname{face}}
\newcommand{\spn}{\operatorname{span}}
\newcommand{\<}{\langle}
\renewcommand{\>}{\rangle}
\newcommand{\Ker}{\operatorname{Ker}}
\newcommand{\Iim}{\operatorname{Im}}
\newcommand{\Tan}{\operatorname{Tan}}
\begin{document}

\title{Geometric and algebraic aspects of spectrality in order unit spaces: a comparison}

    \author{Anna Jen\v cov\'a and
 Sylvia Pulmannov\'{a}{\footnote{ Mathematical Institute, Slovak Academy of
Sciences, \v Stef\'anikova 49, SK-814 73 Bratislava, Slovakia; jenca@mat.savba.sk,
pulmann@mat.savba.sk.  Supported by the grant VEGA 2/0142/20 and  the Slovak Research and Development Agency grant APVV-16-0073.}}}

\date{}

\maketitle

\begin{abstract} Two approaches to spectral theory of order unit spaces are compared: the spectral duality of Alfsen and
Shultz and the spectral compression bases due to Foulis. While the former approach uses the geometric properties of an
order unit space in duality with a base norm space, the latter notion is purely algebraic. It is shown that the Foulis
approach is strictly more general and contains the Alfsen-Shultz approach as a special case. This is demonstrated on two
types of examples: the JB-algebras which are Foulis spectral if and only if they are Rickart and the centrally symmetric state spaces, which may be Foulis spectral while not necessarily Alfsen-Shultz spectral. 

\end{abstract}

\section{Introduction} \label{sc:Intro}

The operational approach to foundations of quantum mechanics works with an
order unit space $A$, in duality with a base normed space $V$, where the distinguished base of the positive cone
represents states of a physical system, whereas elements of the unit interval in $A$ are interpreted as effects, or
dichotomic measurements. This theory goes back to the works of Mackey \cite{mackey1963mathematical} and  Ludwig
\cite{ludwig} and many others, see e.g. \cite{lami2018nonclassical} for a historical account and further
references. More recent approaches include the convex
operational theories  or general probabilistic theories (GPT) (e.g. \cite{barrett2007information,dariano2017quantum}), working mainly
in finite dimension, see also \cite{lami2018nonclassical} for an introduction and overview or 
\cite{wetering2019aneffect} for other approaches.

An important property of any mathematical theory describing quantum mechanics is spectrality, that is, existence of
spectral resolutions of effects (and hence all elements in the space $A$) that allows an integral  expression in terms some
special elements called projections. Perhaps the best well known extension of  spectrality to order unit spaces
 is due to Alfsen and Shultz \cite{AS,AlSh}, motivated by characterization of state spaces of operator algebras and JB-algebras. This theory is based on the notion of
{compressions} on an  order unit space $A$ which employs a separating order and norm duality with a base norm space $V$.  The compressions are in one-to-one
correspondence with special elements of the unit interval in $A$ called {projective units} and also with so-called 
{projective faces} of the set of states.  The duality of $A$ and $V$  is  called {spectral} if there are
''sufficiently many'' compressions on $A$,
it is also assumed that the order unit space is monotone $\sigma$-complete \cite{AS} or that $A=V^*$ \cite{AlSh}. 
Under these conditions, any element has a {spectral resolution} in terms of the projective units and there is a well
defined functional calculus on $A$. 

There are other approaches to spectrality based solely on order theoretic or algebraic properties
of the order unit space, see e.g. \cite{abbati1981aspectral, FPspectres}, or the unit interval in $A$ e.g.
\cite{Gu, Gucomprba, Pucompr},
also more recently \cite{Gucs} or \cite{jencova2019ontheproperties}. The relations of these various definitions are not clear and it is a natural question whether the algebraic approach could lead to the same results as the Alfsen-Shultz theory (if
similar conditions are assumed).

The aim of the present paper is to compare the Alfsen-Shultz approach to that of \cite{FPspectres}. The latter notion of
spectrality follows the works by Foulis \cite{Fcomgroup,Fcompog,  Fgc, Funig} in the more general setting of ordered groups with order unit. This
definition also uses a notion of a compression, but since no duality is used, the definition is different from that of
Alfsen-Shultz. In this case, $A$ is spectral if there are enough compressions to form a {spectral compression base}:
 a set of compressions with given properties. It is proved in \cite{FPspectres} that  this implies existence of spectral
resolutions for elements in $A$.

In order to compare the two theories, we will assume that $A$ is in a separating order and norm duality
with a base norm space $V$ and that the (Foulis) compressions are continuous with respect to this duality, note that we
can always take $V=A^*$ and then continuity always holds.

It is noted in \cite{FPspectres} that the spectral theory of Alfsen-Shultz is a special case of the
spectral theory of Foulis and examples are given showing that the latter notion works also in the case that
the order unit space is not $\sigma$-monotone complete. However, the exact relation of the two theories  is unknown. 
In particular, it is a question whether these two notions of spectrality are equivalent if 
 $A=V^*$ for some base norm space $V$.
We prove that the Foulis spectrality is strictly more general than Alfsen-Shultz spectrality, in the sense that if
$A=V^*$, the Alfsen-Shultz spectrality implies Foulis spectrality, but we show examples that the opposite implication
does not hold. We compare the two notions of compressions and show that these are not the same. 
We also analyze the case of JB-algebras and show that JB-algebras are Foulis spectral if and only if they are Rickart,
 but the Rickart JB-algebras are not  Alfsen-Shultz spectral in general, although the compressions are the same in both cases.

The outline of the paper is as follows. In the next section, we introduce some basic definitions related to order unit
and base norm spaces and dualities between them. In Section \ref{sec:spectrality} we describe the two basic notions of
spectrality that we are going to compare. While the Alfsen-Shultz theory is developed in detail in \cite{AlSh} and is
therefore  introduced only briefly in Section \ref{sec:AS}, the theory proposed in \cite{FPspectres} is much less well
known, with many proofs scattered about several papers and in different settings. For this reason, the exposition of
this theory in Section \ref{sec:FP} is more detailed and quite self-contained, with many results reproved.
 In Section \ref{sec:comparison}, we proceed to comparison of the two theories, here the main result is Theorem
\ref{th:gencomp}. The last two sections contain  examples showing the difference between the two theories. 
In Section \ref{sec:JB}, we study  the Rickart JB-algebras in which case we may have $A\neq V^*$. We work with compressions in Alfsen-Shultz sense and show that they form  a spectral compression base, while Alfsen-Shultz spectrality cannot hold. 
Section \ref{sec:CS} is devoted to  centrally symmetric theories, where $A=V^*$ and the distinguished base in $V$ is
(isomorphic to) the unit ball in a (reflexive) Banach space $(X,\|\cdot\|)$. We show that Alfsen-Shultz  spectrality
holds if and only if the space $X$ is strictly convex and smooth, whereas  smoothness is equivalent to Foulis
spectrality.

\section{Basic notions}

For more details for this section, we refer the reader to \cite{Alf,AlShp1,AlSh}.

Recall that an \emph{order unit space} is an archimedean partially ordered real vector space with a distinguished order
unit. In the sequel, we denote the positive cone by $A^+:=\{a\in A: 0\leq a\}$ and the order unit by $1$, the order unit
space will be denoted by the triple $(A,A^+,1)$. The unit interval
 in $A$ will be denoted by $E:=\{ e\in A: 0\leq e\leq 1\}$. Note that  $E$ endowed  with the partial operation inherited from $+$ in $A$  is an effect algebra, 
 see \cite{FoBe} for  the  definition and \cite{DvPu} for more information on effect algebras and related structures.

The order unit norm $\|.\|_1$ on $A$ is defined by  
\[
\|a\|_1=\inf \{\lambda\in {\mathbb R}^+: -\lambda \leq a \leq \lambda \},\qquad a\in A.
\]

A \emph{base norm space} is a partially ordered normed real vector space $V$ with a generating cone $V^+$ and a
distinguished base $K$  of $V^+$ such that $K$ lies in a hyperplane $H\not\ni 0$ in $V$ and $co(K\cup -K)$ is the unit ball in $V$. The base norm space will be denoted
by $(V,K)$. The norm of $V$ is the 
base norm with respect to $K$ denoted by $\|\cdot\|_K$ and is given by
\[
\|v\|_K=\inf\{\lambda+\mu: v=\lambda x-\mu y,\ \lambda,\mu\ge 0, x,y\in K\},\qquad v\in V. 
\]

An order unit space $(A,A^+,1)$ and a base norm space $(V,K)$ are in \emph{separating order and norm duality} 
if there is a duality $\<\cdot,\cdot\>: A\times V\to  \mathbb R$  such that
\begin{enumerate}
\item[(i)] $ a\in A^+ \ \Leftrightarrow\ \<a,\varphi\> \geq 0$  for all $\varphi\in V^+$,

\item[(ii)] $ \varphi\in V^+ \ \Leftrightarrow \ \<a,\varphi\> \geq 0$ for all $a\in A^+$,

\item[(iii)] for all $a\in A$ and $\varphi\in V$:
\[
\|a\|_1= \sup_{\|\sigma\|_K\le 1}|\<a,\sigma\>|,\qquad \|\varphi\|_K=\sup_{\|a\|_1\le 1}|\<a,\varphi\>|.
\]
\end{enumerate}
In this case, $\|\varphi\|_K=\<1,\varphi\>$ for $\varphi\in V^+$ and $K=\{\rho\in V^+:\ \<1,\rho\>=1\}$.

\begin{example}\label{ex:duality}
Let  $(A,A^+,1)$ be an order unit space and let $V:=A^*$, $V^+:=\{\varphi\in V,\ \<a,\varphi\>\ge 0\ \forall a\in A^+\}$ and
$K:=\{\rho\in V^+:\ \<1,\rho\>=1\}$. Then $A$ and $V$ are in separating order and norm duality.

Similarly, let $(V,K)$ be a base norm space and let $A:=V^*$, $A^+:=\{a\in A,\ \<a,\varphi\>\ge 0,\ \forall \varphi\in
V^+\}$ and $1\in A^+$ be the functional determining the hyperplane $H$ containing $K$. Then $A$ and $V$ are in
separating order and norm duality.  

\end{example}

\begin{example}\label{ex:vNC*} Let $A=\mathcal A^{sa}$ be the self-adjoint part of the C*-algebra $\mathcal A$. Let 
 $A^+=\mathcal A^+=\{b^*b,\ b\in \mathcal A\}$ be the positive cone in $\mathcal A$ and let $1$ be the identity in
$\mathcal A$. Then $(A,A^+,1)$ is an order unit space. If $V=\mathcal A^*$ and $K=\mathfrak S(\mathcal A)$ is the set of
states on $\mathcal A$, then $(V,K)$ is a base norm space in separating order and norm duality with $A$.
If $\mathcal A$ is a von Neumann algebra, then the predual $V=\mathcal A_*$ with the set $K=\mathfrak S_*(\mathcal A)$ of
normal states  is a base norm space in separating order and norm duality with $A$.

More generally, we may take $A$ to be a JB-algebra resp. a JBW-algebra and obtain similar results.

\end{example}


Let $(A,A^+,1)$ and $(V,K)$  be an order unit space and a base norm space in separating order and norm duality. We
denote by   $\sigma(A,V)$ be the locally convex topology in $A$ or $V$, given by the duality of $A$ and $V$. Unless
otherwise stated, we will always consider $A$ and $V$ endowed with these topologies.

Recall that a \emph{face} of the positive cone $A^+$ (or similarly of $V^+$) is a hereditary subcone $F\subseteq A^+$:
$ a\in F$ and $b\le a$ implies $b\in F$. The face of $A^+$ generated by $a\in A^+$ will be denoted as
\[
\face_{A^+}(a):=\{ b\in A^+: b\le \lambda a,\ \lambda\ge 0\}. 
\]
A subcone $F\subseteq V^+$ is a face of $V^+$ if and only if $F\cap K$ is a face of $K$: $\rho\in F\cap K$ and $\rho=\lambda
\sigma_1+ (1-\lambda) \sigma_2$ with $\lambda\in (0,1)$ and $\sigma_1,\sigma_2 \in K$ implies  $\sigma_1,\sigma_2\in
F\cap K$.

Let $C\subseteq A$, $W\subseteq V$ be any subsets. We will use the notations
\[
C^\circ:=\{\varphi\in V:\ \<a,\varphi\>= 0,\ \forall a\in C\}, \quad W^\circ:=\{a\in A:\ \<a,\varphi\>= 0,\ \forall
\varphi\in W\}.
\]
Note that $C^{\circ\circ}$ is the closed linear span of $C\cup\{0\}$, similarly for $W^{\circ\circ}$.

If $C\subseteq A^+$ and $W\subseteq V^+$, we will denote 
\[
C^\bullet=\{\varphi\in V^+:\ \<a,\varphi\>= 0,\ \forall a\in C\}, \quad W^\bullet:=\{a\in A^+:\ \<a,\varphi\>= 0,\ \forall
\varphi\in W\}.
\]
Note that $C^\bullet$ is a face of $V^+$, similarly for $W^\bullet$. A face  $F\subseteq V^+$ is called \emph{exposed} 
 if there is some $a\in A^+$ such that $F=\{a\}^\bullet$ and \emph{semiexposed} if $F=C^\bullet$ for some $C\subseteq
A^+$, equivalently, $F^{\bullet\bullet}=F$. We have similar definitions for faces of $V^+$. Note that a face $F\subseteq V^+$
is semiexposed iff $F\cap K$ is semiexposed, which means that there is a subset of effects $D\subseteq E$
 such that 
\[
F\cap K=\{\rho\in K: \ \<a,\rho\>=1,\ \forall a\in D\},
\]
and $F$ is exposed iff we may take a singleton $D=\{a\}$.

\section{Spectrality in order unit spaces}\label{sec:spectrality}

 In this section, we will recall the definitions and basic results of the two
approaches to spectrality in order unit spaces that we are going to compare. Throughout the rest of the
 paper, we will work with a pair  $(A,V)$ of an order unit space $(A,A^+,1)$ and a base norm space $(V, K)$ in separating order and norm duality.

\subsection{Spectral duality of Alfsen and Shultz}\label{sec:AS}

We first introduce the theory of spectral duality due to Alfsen and Shultz. We refer to 
 \cite{AS, AlSh} for details. 

Let $J:A\to A$ be a positive idempotent linear map, continuous in the $\sigma(A,V)$-topology. We will say that $J$ is \emph{normalized} if $\|J\|\le 1$,
equivalently $p:=J(1)\le 1$. Let us denote  
\begin{equation}\label{eq:kerim}
\Ker^+ J :=\{a\in A^+: \ J(a)=0\}\qquad \Iim^+ J :=\{a\in A^+: \ J(a)=a\}.
\end{equation}
We say that two such maps  $J$ and $J'$ are \emph{complementary}  if
\[
\Ker^+J=\Iim^+J'\quad \mbox{ and }\quad   \Ker^+J'=\Iim^+J.
\]
In this case we say that $J'$ is a \emph{complement} of $J$.

Notice that $J$ has an  adjoint map $J^*:V\to V$, which is positive and idempotent as well and we have similar definitions of 
 $\Ker^+(J^*)$, $\Iim^+(J^*)$ and complementarity.  We will say that  the maps  $J$ and $J'$ are \emph{bicomplementary} 
if  they are complementary and $(J')^*$ is a complement of $J^*$. In this case, we say that $J$ (or $J'$) is 
\emph{bicomplemented}. Note that then $J$ has a unique complement $J'$ \cite[Corollary 7.11]{AlSh}.

\begin{definition}\label{de:compr} A \emph{compression} on $A$ is a bicomplemented normalized $\sigma(A,V)$-continuous positive idempotent linear map
 $A\to A$.  
\end{definition}

If $J$ is a compression,  the element $p=J(1)$ is called a \emph{projective unit}. If $p$ is a projective unit, then
there is a unique compression  $J$ such that $J(1)=p$ and then $J'(1)=1-p$ for the complementary compression. This
implies that any compression has a unique complement.

\begin{definition}\label{def:spectral}\cite[Def. 8.42]{AlSh} We say that the spaces $A$ and $V$ are in \emph{spectral duality} if $A=V^*$ and 
for every $a\in A$, there is a least compression $J$ such that $J(a)\ge a$ and $J(a)\ge 0$.

\end{definition}

Note that if $a\in A$ and $J$ is a compression as in the above definition with complement $J'$, then we have
\[
a=J(a)+J'(a)=a^+-a^-,\qquad a^+:=J(a),\ a^-:=-J'(a)\ge 0.
\] 
In general, for  $a\in A$, $b,c\in A^+$, we say that $a=b-c$ is an \emph{orthogonal decomposition} of $a$ if there is a
compression $J$ such that $J(b)=b$, $J(c)=0$.

There are several equivalent characterizations of spectral duality. Here an important role is played by 
 the following assumption:

\begin{assumption}\label{ass:SH}\cite{AlSh}
 Every exposed face $F\subseteq K$ is projective, which means that it is determined by a projective unit. 
   Since $K$ is a base of $V^+$, this means that for every $a\in A^+$ there is a projective unit $p$ such that 
\[
\{a\}^\bullet=\{p\}^\bullet.
\]

\end{assumption}

By \cite[Thms. 8.52 and 8.55]{AlSh}, $A$ and $V$ are in spectral duality iff the Standing Hypothesis 
 holds and every element $a\in A$ has a unique orthogonal decomposition: $a=a^+-a^-$.

The importance of the notion of spectral duality is the fact that it allows
us to define spectral resolutions of elements in $A$, as we briefly describe next. For $a\in A$, let 
\[
e_\lambda:=1-J_{\lambda,a}(1),
\]
 where $J_{\lambda,a}$ is the compression corresponding to $a-\lambda$ (by Definition \ref{def:spectral}). The family 
$\{e_\lambda\}_{\lambda\in \mathbb R}$ has exactly the properties of the spectral resolution of a self-adjoint element
in a von Neumann algebra. In particular, we have
\[
a=\int\lambda de_\lambda,
\]
 where the integral is defined as the norm-limit of approximating Riemann-Stieltjes sums. The family is uniquely  determined by $a$ and called the \emph{spectral resolution} of $a$.

The following is a prototypical example of a spectral duality.

\begin{example}\label{ex:vN_JBW} Let $A=\mathcal M^{sa}$ be the self-adjoint part of a von Neumann algebra $\mathcal M$.
For any projection $p\in A$, the map 
\[
U_p: a\mapsto pap,\qquad a\in A
\]
is a compression on $A$ and $p=U_p(1)$. Moreover, any compression on $A$ is
necessarily of this form for some projection $p$, \cite[Theorem 7.23]{AlSh}. In this way, projective units generalize the projections in von Neumann
algebras. Furthermore, $A=V^*$ where $V$ is the hermitian part of the predual of $\mathcal M$ and $A$ and $V$ are in
spectral duality.  For $a\in A$, let 
 $a=a^+-a^-$ be the decomposition into the positive and negative part and
 let $p$ be the support projection of $a^+$, then $U_p(a)=a^+\ge 0$,  $a\le a^+=U_p(a)$ and $U_p$ is the compression 
 corresponding to $a$ in Definition \ref{def:spectral}. 
It is easily seen that in this case, $\{e_\lambda\}$ for $a\in A$ is the usual  spectral
resolution for $a$. Similar results hold in the case of a general JBW algebra.

\end{example}

\subsection{Compression bases and spectrality}\label{sec:FP}

In this section, we discuss the notion of spectrality in an order unit space along the lines of \cite{FPspectres}.
This notion is based on the works \cite{Fcomgroup, Fcompog, Fgc, Funig} on compressions on unital ordered abelian groups, 
that were applied to order unit spaces in \cite{FPspectres}.
We stress that the original definitions  are of algebraic nature and no duality with a base norm space is
assumed. Since we want to compare this to the theory  of Alfsen and Shultz, we have to admit such duality here and the maps  are assumed continuous. Note
that we can always put $V=A^*$, then all the involved mappings are automatically continuous (with respect to the duality 
 of $A$ and $V$). To distinguish the maps used here from the
compressions in the previous paragraph,  we use the term 
F-compressions.

The following definition is based on \cite[Definition 1.1 and Lemma 1.1]{FPspectres}. 
\begin{definition}\label{de:Fcompr}  Let $J:A\to A$ be a  $\sigma(A,V)$-continuous positive 
linear mapping. Then $J$ is an  \emph{F-compression with focus $p$} on $A$ if for all  $e\in E$,
\begin{enumerate}
\item[(F1)] $J(1)=p\in E$ (that is, $J$ is normalized),
\item[(F2)] $e\leq p \ \implies \ J(e)=e$,
\item[(F3)] $J(e)=0\ \implies \ e\leq 1-p$.
\end{enumerate}
If $J$ satisfies (F1) and (F2) but not necessarily (F3), we say that $J$ is a \emph{retraction}.

\end{definition}

It is easily seen from this definition that any retraction is idempotent.
Let $J$ be an F-compression and let $\Ker^+J$ and $\Iim^+ J$ be as in \eqref{eq:kerim}. The following equalities follow easily from the properties (F2) and (F3):
\begin{align}
\Ker^+(J)&=\{a\in A^+,\ \exists t>0, a\le t(1-p)\}=\face_{A^+}(1-p)\label{eq:kerj}\\
\Iim^+(J)&=\face_{A^+}(p),\label{eq:imj}
\end{align}
 We next collect some easy observations on F-compressions and their foci for later use.

\begin{lemma}\label{le:complementary} Let $J$ and $J'$ be F-compressions, $J(1)=p$ and $J'(1)=p'$. Then  $J$ and $J'$ are
complementary if and only if  $p'=1-p$.
\end{lemma}

\begin{proof}
Let $J$ and $J'$ be complementary. Then
 $1-p\in \Ker^+J=\Iim^+ J'$ and $p\in \Iim^+ J= \Ker^+J'$, so that $J'(p)=0$ and
$1-p=J'(1-p)=J'(1)=p'$. 
The converse is clear from the equalities \eqref{eq:kerj} and \eqref{eq:imj}.
\end{proof}

\begin{lemma}\label{lemma:principal} Let $p$ be the focus of a retraction $J$. Then $p$ is 
 a \emph{principal element} in $E$: for any $a,b\in E$ such
that $a+b\le 1$ and $a,b\le p$ we have $a+b\le p$. 
This property is equivalent to  the following \emph{facial property} (cf. \cite[Def. 8.45]{AlSh}):
\[
\face_{A^+}(p)\cap E=[0,p].
\]
If $J$ is an F-compression, the same is true for $1-p$.
\end{lemma}

\begin{proof} Let $a,b\in [0,p]$ be such that $a+b\le 1$. Then  by property (F2) we have
$a=J(a)$, $b=J(b)$, so that
\[
a+b=J(a+b)\le J(1)=p.
\]
Next, assume that an element $q\in E$ has the facial property and let $a,b\le q$, $a+b\in E$. Then 
$a+b\le 2q$, so that $a+b\in \face_{A^+}(q)\cap E=[0,q]$. Conversely, assume that $q$ is principal, and let 
$0\leq a\leq \lambda q$ for some $\lambda>0$.  If $\lambda \le 1$ there is nothing to prove, so assume that $\lambda>1$.
 Then $(1/ \lambda) a\leq q$ and  $a= (1/\lambda)a + (1-1/\lambda )a$. We can decompose 
\[
1-1/\lambda=\sum_i\beta_i, \qquad i=1,\ldots
,n,\ 0\le \beta_i\leq 1/\lambda,\ \forall i.
\]
Then $\beta_i a\leq (1/\lambda)a\leq q$ for all $i$ and since $q$ is principal,  we must have $a=(1/\lambda)a+ \sum_i \beta_ia\leq q$.
The last statement follows easily by \eqref{eq:kerj} and the property (F3).

\end{proof}

\begin{lemma}\label{lemma:princ_ext_sharp} For $p\in E$, consider the following statements:
\begin{enumerate}
\item[(a)] $p$ is principal;
\item[(b)] $p$ is extremal in $E$;
\item[(c)] $p$ is \emph{sharp}: if $a\in E$ is such that $a\le q$ and $a\le 1-q$, then $a=0$.
\end{enumerate}
Then (a) $\implies$ (b) $\implies$ (c).
\end{lemma}

\begin{proof} Assume (a) and let $a,b\in E$, $\lambda\in (0,1)$ be such that 
$\lambda a+(1-\lambda)b=p$. By the facial property of $p$, we must have $a,b\le p$, but this is possible only if
$a=b=p$, so that $p$ is extremal in $E$. Assume (b) and let $a\in E$, $a\le p$, $a\le 1-p$. Then $p\pm a\in E$ and 
 $p=\tfrac12((p-a)+(p+a))$, so that $p-a=p+a=p$ and $a=0$. Hence $p$ is sharp.

\end{proof}

We say that the elements $a,b\in E$ are \emph{Mackey compatible}, if  there are 
some elements $c,a_1,b_1\in E$ such that $c+a_1+b_1\le 1$ and $a=c+a_1$, $b=c+b_1$.
A sub-effect algebra $P$ in $E$ is
\emph{normal} if for all $d,e,f\in E$ such that $d+e+f\leq 1$ and $d+e,d+f
\in P$ we have $d\in P$ \cite[Definition 1]{Funig}. Note that this implies that elements of $P$ are compatible in $E$
iff they are compatible in $P$.

\begin{definition}\label {de:comprba} {\rm \cite[Definition 2]{Fgc}} A \emph{compression base} for $A$ is a family 
$(J_p)_{p\in P}$ of F- compressions on $A$, indexed by their own foci, such that $P$ is a normal subalgebra of $E$ and
whenever $p,q,r\in P$ and $p+q+r\leq 1$, then $J_{p+r}\circ J_{q+r}=J_r$. Elements of $P$ will be called
\emph{projections}.
\end{definition}

\begin{example}\label{ex:CXR} Let $X$ be a compact Hausdorff space and let $A=C(X,\mathbb R)$ be the set of continuous
functions $X\to \mathbb R$. With the usual ordering of functions and $1=1_X$ the constant unit, $A$ is an
order unit space and the order unit norm coincides with the maximum norm on $C(X,\mathbb R)$. Since $A$ is the
self-adjoint part of the C*-algebra $C(X,\mathbb C)$, we know by \cite{Funig} (see also Theorem \ref{th:compressible}
below) that the retractions on $A$ are precisely
of the form $U_p(f)=pf$, $f\in A$, where $p\in A$ is a projection, which means that $p$ is the characteristic function
of a clopen subset of $X$. By a similar reasoning as in the proof of Corollary \ref{coro:JBcompb}, we have that 
 $(U_p)_{p\in \mathcal P(X)}$ is a compression base, here  $\mathcal P(X)$ is the set of all projections in $A$. 

We will encounter this example repeatedly below and we will always assume that $C(X,\mathbb R)$ is endowed with the
compression base $\{U_p\}_{p\in \mathcal P(X)}$.

\end{example}

In the sequel, we fix a compression base $(J_p)_{p\in P}$ for $A$. Note that since $P$ is a subalgebra, any $J_p$ has a
 (fixed) complementary F-compression $J_{1-p}$. Observe also that using the fact that all projections are principal elements, 
it is easily seen  that if $p,q\in P$ and $p+q\le 1$, then $p+q=p\vee q$ in $P$, so that $P$ with the orthocomplementation 
$p\mapsto 1-p$ is an \emph{orthomodular poset} (OMP) \cite{PtPu, Harding}. It follows that if $p,q\in P$ are Mackey 
compatible, then 
 $p\wedge q$ and $p\vee q$ exists in $P$. It can be shown as in  \cite[Thm. 2.5]{Fgc} that $P$ is a \emph{regular} OMP, 
 which means that any pairwise Mackey compatible subset in $P$ is contained in a Boolean subalgebra of $P$
(\cite{Harding}). A maximal subset of pairwise Mackey compatible elements is
called a \emph{block} of $P$. Every element $p\in P$ is contained in some block $B$ of $P$.

\subsubsection{Compatibility}

The following notion  for $p\in P$ and $a\in A$ was introduced in \cite[Def. 1.5]{FPspectres}, see also
\cite[Def. 7.41]{AlSh} for projective units $p$.

\begin{definition}\label{def:compatibility} For $p\in P$ and $a\in A$, we say that $a$ is  \emph{compatible with $p$} if 
\[
a=J_p(a)+J_{1-p}(a).
\] 
The set of all $a$ compatible with $p\in P$ will be denoted by $C(p)$. For a subset $Q\subseteq P$, we denote 
 $C(Q):=\bigcap_{p\in Q}C(p)$. 
If $B\subseteq P$ is a block of $P$, the set $C(B)$ is called a \emph{C-block} of $A$.

\end{definition}

The following facts are easily checked (see also \cite[Lemma 1.3]{FPspectres}).

\begin{lemma}\label{lemma:compatible}
Let $a\in A$, $p\in P$. Then
\begin{enumerate}
\item[(i)] If $J_p(a)\le a$, then $a\in C(p)$.
\item[(ii)] If $a\in A^+$, then $a\in C(p)$ if and only if $J_p(a)\le a$.
\item[(iii)] If $a\in E$, then $a\in C(p)$ if and only if $a$ and $p$ are {Mackey compatible}.
\item[(iv)] If $q\in P$, then $q\in C(p)$ if and only if $p\in C(q)$ if and only if 
\[
J_pJ_q=J_qJ_p=J_{p\wedge q}.
\]
\end{enumerate}

\end{lemma}

 We denote the set of all projections compatible with $a\in A$ by $PC(a)$
 and for $B\subseteq A$, $PC(B):=\bigcap_{a\in B} PC(a)$.

\begin{lemma}\label{lemma:pca}
For $B\subseteq  A$ we have:
\begin{enumerate}
\item[(i)] If $p_1,\dots,p_n\in PC(B)$ are such that $\sum_i p_i\le 1$, then
\[
J_{\sum_i p_i}(a)=\sum_iJ_{p_i}(a)\qquad \forall a\in B.
\] 
\item[(ii)] For compatible $p,q\in PC(B)$, we have $p\vee q, p\wedge q\in PC(B)$.
 \end{enumerate}
Consequently, $PC(B)$ is a normal subalgebra in $E$ and a regular OMP.

\end{lemma}
 
\begin{proof} 
Let $p,q\in PC(a)$, $p+q\le 1$ and let $r=1-p-q$. Then since $p\le p+q$ and $1-p=r+q$, we have for $a\in B$
\[
J_{p+q}(a)=J_{p+q}(J_p(a)+J_{1-p}(a))=J_p(a)+J_{p+q}J_{r+q}(a)=J_p(a)+J_q(a),
\]
 By induction, this proves (i). 

Assume that  $p,q\in PC(B)$ are compatible and put $p'=1-p$, $q'=1-q$. Then for all $a\in B$
\begin{align*}
a&=J_p(a)+J_{p'}(a)=J_p(J_q(a)+J_{q'}(a))+J_{p'}(J_q(a)+J_{q'}(a))\\
&= J_{p\wedge q}(a)+J_{p\wedge q'}(a)+ J_{p'\wedge
q}(a)+J_{p'\wedge q'}(a).
\end{align*}
Since $1-p\wedge q=p\wedge q'+q'\wedge p+q'\wedge p'$, all the involved elements are compatible and we obtain by Lemma
\ref{lemma:compatible}
\begin{align*}
J_{1-p\wedge q}(a)&= J_{1-p\wedge q}(J_{p\wedge q'}(a)+ J_{p'\wedge
q}(a)+J_{p'\wedge q'}(a))\\
&=J_{p\wedge q'}(a)+ J_{p'\wedge
q}(a)+J_{p'\wedge q'}(a)
\end{align*}

so that 
\[
a= J_{p\wedge q}(a)+J_{1-p\wedge q}(a).
\]
Hence $p\wedge q\in PC(B)$, (ii) follows from $p\vee q=1-p'\wedge q'$.  The last statement follows by \cite[Cor.
2.1.14]{Harding}.

\end{proof}

\begin{lemma}\label{lemma:CQ_subspace} For $Q\subseteq P$, $C=C(Q)$ is a norm-closed subspace and an order unit space,
with order unit 1. Let $\tilde J_p:=J_p|_C$, then $\{\tilde J_p\}_{p\in P\cap C}$ is a compression base in $C$. 
If $Q$ is a block, then $P\cap C=Q$. 

\end{lemma}

\begin{proof} Let $q\in P$, then $C(q)=\{a\in A, a=J_q(a)+J_{1-q}(a)\}=J^{-1}(0)$, where $J=id-J_q-J_{1-q}$ is a 
bounded linear map. Hence $C(q)$ is a norm-closed subspace and since $1\in C(q)$, it is an order unit space with order
unit 1. Further, let $p\in C(q)\cap P$, then since $J_pJ_q=J_qJ_p$, we have for  any $a\in C(q)$
\[
J_p(a)=J_p(J_q(a)+J_{1-q}(a))=J_q(J_p(a))+J_{1-q}(J_p(a)),
\]
so that $J_p(C(q))\subseteq C(q)$. Since $C=\cap_{q\in Q} C(q)$, we see that $C$ is a norm closed subspace in $A$ and an
order unit space with order unit 1. Moreover, for $p\in C\cap P$,  $\tilde J_p(C)\subseteq C$ and it is straightforward
that $\tilde J_p$ is an F-compression. By Lemma \ref{lemma:pca}, $P\cap C(Q)= PC(Q)$ is a normal subalgebra in $E$ and
hence also in $E\cap C$. It is also easily checked that $\{\tilde J_p\}_{p\in C\cap P}$ is a compression
base in $C$. 

If $Q$ is a block, then clearly $Q\subseteq C(Q)\cap P$, the other inclusion follows by maximality of $Q$.

\end{proof}

%
%
%
%
%
%
%

The following subset  can be seen as the \emph{bicommutant} of $a$ in $P$: 
\[
P(a):=PC(PC(a)\cup\{a\}),
\]
i.e. the set of all projections compatible with $a$ and with all projections compatible with $a$. 
Note that all elements in $P(a)$ are pairwise compatible, so that by Lemma \ref{lemma:pca}, $P(a)$ is a boolean
subalgebra in $P$. If $q\in P$, then $P(q)=\{0,1,q,1-q\}$.

\subsubsection{Projection cover property}

\begin{definition}
We say that the compression base $(J_p)_{p\in P}$ (or $A$)
  has the \emph{projection cover property} if for every effect $e\in E$ there is an element $p\in P$ 
such that for $q\in P$, we have $e\le q$ iff $p\le q$. Such an element $p$ is necessarily unique and is called the
 \emph{projection cover} of $e$,  denoted by $e^0$.
\end{definition}

In this section, we assume that $(J_p)_{p\in P}$ has the projection cover property.

\begin{lemma}\label{lemma:pc} Assume that $(J_p)_{p\in P}$ has the projection cover property.
For $a\in E$, $a^0\in P(a)$.

\end{lemma}

\begin{proof} Let $q\in PC(a)$, then $a=b+c$, with $b=J_q(a)$, $c=J_{1-q}(a)$. Since $b\le q$, $c\le 1-q$, we have $b^0\le q$ and $c^0\le 1-q$, so that 
$b^0,c^0\in PC(q)$. We will show that  $a^0=b^0+c^0=b^0\vee c^0\in PC(q)$. Indeed, since $a=b+c\le b^0+c^0$, we have $a^0\le b^0+c^0$. On the
other hand, $b,c\le a\le a^0$, so that $b^0,c^0\le a^0$ and $b^0+c^0=b^0\vee c^0\le a^0$.  
The fact that $a^0\in PC(a)$ is clear from $a\le a^0$.
 
\end{proof}

The following theorem was proved in \cite[Theorem 6.4 (iii)]{Fgc} in a more general context. We give a simpler proof
here.

\begin{theorem}\label{th:projcovoml}
  Suppose that $(J_p)_{p\in P}$ has the projection cover property. Then
\begin{enumerate}
\item[(i)] $P$ is an orthomodular lattice (OML);
\item[(ii)] $P$ is sup/inf closed in $E$, in the sense that for every subset $M\subseteq P$, whenever the supremum $\vee
M$ exists in $E$, then $\vee M\in P$, and similarly for the infimum $\wedge M$.

\end{enumerate}

\end{theorem}

\begin{proof} Since, as noted above, $P$ is an OMP, we only need to prove that it is a lattice, in
fact, it is enough to show that any two elements $p,q\in P$ have a supremum in $P$. Let us denote
$r_{\lambda}:=(\lambda p+ (1-\lambda) q)^0$ for some $\lambda\in (0,1)$. Then since $\lambda p\le r_\lambda$, we
have $p\le r_\lambda$ by the facial property, similarly $q\le r_\lambda$. Let $r\in P$ be such that $p,q\le r$, then 
$\lambda p+(1-\lambda)q\le r$, so that $r_\lambda\le r$ (since $r_\lambda$ is the projection cover). Hence we may put 
$p\vee q=r_\lambda$, for any $\lambda\in (0,1)$. 

Let $M\subseteq {P}$ and $\wedge M=:a$ exists in $E$. For all $b\in M$, $a\leq b \, \implies \, a^0\leq b$, hence $a^0\leq \wedge M =a$, which entails $a=a^0\in {P}$.

\end{proof}

\subsubsection{Comparability}

\begin{definition}\label{de:compar} {\rm \cite[Def. 1.6]{FPspectres}} We say that the compression base
$(J_p)_{p\in P}$ in  $A$ has
the \emph{comparability property} if, for every $a\in A$, $P^{\pm}(a)\neq \emptyset$, where 
\[
P^{\pm}(a):=\{ p\in P(a)\mbox{ and} \ J_{1-p}(a)\leq 0\leq J_p(a)\}.
\]
\end{definition}

Let  $p\in P^\pm(a)$ and put $b:=J_{p}(a)$, $c:=-J_{1-p}(a)$, then we have 
\begin{equation}\label{eq:Port_dec}
a=b-c,\quad b,c \in A^+, \quad J_p(b)=b,\ J_p(c)=0.
\end{equation}
 Any decomposition  of $a$  of the form \eqref{eq:Port_dec}  for some $p\in P$ is called  a
$P$-\emph{orthogonal decomposition} of $a$. 

\begin{prop}\label{prop:orthogonal} If $P^\pm(a)\ne \emptyset$, then $a\in A$ has a unique $P$-orthogonal decomposition,
 which will be written as
\[
a=a^+-a^-.
\]
Moreover we have $a^+,a^-, |a|:=a^++a^-\in C(PC(a))$. 
\end{prop}

\begin{proof} Let $p\in P^{\pm}(a)$, we have seen that $a$ has a $P$-orthogonal decomposition $a=J_p(a)-(-J_{1-p}(a))$.
To prove uniqueness, let $a=b-c$ with $b,c\in A^+$ and $J_q(b)=b$, $J_q(c)=0$ for some $q\in P$. Then 
  $a\in C(q)$ and hence $p$ and $q$ are compatible, by definition of $P^\pm(a)$. Since $J_{1-p}(a)\le 0$ and $J_q$ is
positive, we have $J_qJ_{1-p}(a)\le 0$, similarly, we also have $J_{1-p}J_q(a)\ge 0$. But $q$ and $1-p$ are compatible,
so that 
\[
0\le J_{1-p}J_q(a)=J_qJ_{1-p}(a)\le 0,
\]
hence $J_{1-p}J_q(a)=0$ and $J_pJ_{1-q}(a)=0$ follows by a similar argument. We obtain
\[
J_p(a)=J_pJ_q(a)=J_qJ_p(a)=J_q(a),
\]
this proves that the $P$-orthogonal decomposition is unique. For the last statement, let $q\in PC(a)$, then 
\[
a^+=J_p(a)=J_p(J_q(a)+J_{1-q}(a))=J_q(a^+)+J_{1-q}(a^+),
\]
so that $a^+\in C(PC(a))$. The proof for $a^-$ and $|a|$ is similar.

\end{proof}

We will assume below that $(J_p)_{p\in P}$ is a compression base with the comparability property. We will  show that
in this case any $a\in A$ is contained in some C-block of $A$ and that the C-blocks are isomorphic to (sub)spaces of
functions as in Example \ref{ex:CXR}. Moreover,  we will  extend the notion of compatibility to all pairs of elements in
$A$. We start by showing that in this case all sharp elements are projections (cf. \cite{Fgc, Pucompr}).

\begin{lemma}\label{le:principal} Suppose that $(J_p)_{p\in P}$ has the comparability property and let $q\in E$. Then 
$q\in P$ if and only if $q$ is sharp.
\end{lemma}

\begin{proof}
We know by Lemmas \ref{lemma:principal} and \ref{lemma:princ_ext_sharp} that all elements in $P$ are sharp. Conversely, let $p\in E$ be sharp
 and let $a=2p-1$. Let $q\in P^\pm(a)$, then $J_{q}(a)\ge 0$, so that $J_{q}(1-p)\le J_{q}(p)$. On the other hand,
 since $a\in C(q)$, it is easily seen that  $q$ is compatible with $p$, so that 
$J_{q}(p)\le p$ and also  $J_{q}(1-p)\le 1-p$.  Hence $J_{q}(1-p)\le p,1-p$ and we must have $J_{q}(1-p)=0$. It follows
that
\[
a^+=J_q(a)=J_q(1-2(1-p))=q.
\]
Similarly, we obtain that $a^-=1-q$, so that $a=a^+-a^-=2q-1$ and $p=q\in P$.

\end{proof}

%
%
%
%
%

\begin{lemma}\label{lemma:descending}(cf. \cite[Thm. 3.7]{Fgc}) Assume that $(J_p)_{p\in P}$ has the comparability property. Let $a\in A$
and let $\lambda_1\le\lambda_2\le\dots$ be a sequence. Then there are projections  $q_i\in P^\pm(a-\lambda_i)$ such that
 $q_{i}\ge q_{i+1}$, $i=1,2,\dots$.

\end{lemma}

\begin{proof} We take any  $q_1\in P^\pm(a-\lambda_1 )$ and construct the rest of the sequence by induction. 
So assume that for
some $k\ge 1$ we already have a sequence $q_1\ge \dots \ge q_k$ with the required properties. Choose some $q\in
P^\pm(a-\lambda_{k+1})$ and put $q_{k+1}:=q\wedge q_k$. Clearly, $q_{k+1}\in P(a)=P(a-\lambda_{k+1})$ and
\[
J_{q_{k+1}}(a-\lambda_{k+1})=J_{q_k}J_q(a-\lambda_{k+1})\ge 0.
\]
Let $q'=1-q$, $q_k'=1-q_k$. Since $q,q_k$ are elements of the boolean algebra $P(a)=P(a-\lambda_{k+1})$, 
so are $q',q'_k$ and we have
$1-q\wedge q_k=q'\wedge q_k+q\wedge q_k'+q'\wedge q_k'$. By Lemma \ref{lemma:pca}, 
\[
J_{1-q\wedge q_k}(a-\lambda_{k+1})= (J_{q'\wedge q_k}+J_{q\wedge q_k'}+J_{q'\wedge q'_k})(a-\lambda_{k+1})
\]
and we have 
\begin{align*}
J_{q'\wedge q_k}(a-\lambda_{k+1} )&= J_{q_k}J_{q'}(a-\lambda_{k+1} )\le 0,\\
J_{q\wedge q_k'}(a-\lambda_{k+1} )&\le J_{q\wedge q_k'}(a-\lambda_k )=J_qJ_{q_k'}(a-\lambda_k )\le 0\\
J_{q'\wedge q_k'}(a-\lambda_{k+1} )&\le J_{q'}J_{q_k'}(a-\lambda_k )\le 0.
\end{align*}
Hence $J_{1-q_{k+1}}(a-\lambda_{k+1}1)\le 0$ and so $q_{k+1}\in P^\pm (a-\lambda_{k+1}1)$.

\end{proof}

\begin{theorem}\label{thm:span} Assume that $(J_p)_{p\in P}$ has the comparability property. Then any $a\in A$ is in the
norm-closed linear span of $P(a)$: $a\in \overline{\spn}(P(a))$, the same is true for $a^+,a^-,|a|$.

\end{theorem}

\begin{proof} For $\epsilon >0$, let us choose a sequence 
\[
-\|a\|_1=:\lambda_0< \lambda_1<\dots <\lambda_n<\lambda_{n+1}:=\|a\|_1,\quad \lambda_{i+1}-\lambda_{i}\le \epsilon, \
i=0\dots,n.
\] 
By Lemma \ref{lemma:descending}, there is a sequence $q_i\in P^\pm(a-\lambda_i)$, $i=0,\dots n+1$ such that 
 $q_i\ge q_{i+1}$, note that we may put $q_0:=1$ and $q_{n+1}:=0$. Put
\[
p_i:= q_{i-1}-q_i=q_{i-1}\wedge q_i',\quad i=1,\dots, n+1.
\] 
Then $p_i\in P(a)$, $\sum_i p_i=1$, so by Lemma \ref{lemma:pca}, $a=\sum_i J_{p_i}(a)$. We also have
\[
\lambda_{i-1}p_i\le J_{p_i}(a)\le \lambda_i p_i,\qquad i=1,\dots, n+1
\]
Indeed, $J_{p_i}(a-\lambda_i)=J_{q_{i-1}}J_{q_i'}(a-\lambda_i)\le 0$, so that $J_{p_i}(a)\le \lambda_ip_i$. The other
inequality follows from
$J_{p_i}(a-\lambda_{i-1})=J_{q_i'}J_{q_{i-1}}(a-\lambda_{i-1})\ge 0$. 
For any choice of $\xi_i\in (\lambda_{i-1},\lambda_i)$, we obtain
\[
-\epsilon p_i\le J_{p_i}(a)-\xi_ip_i\le \epsilon p_i,\qquad i=1,\dots, n+1,
\]
so that, summing up over $i$, we obtain $-\epsilon \le a-\sum_i \xi_ip_i\le \epsilon$ which means that
$\|a-\sum_i\xi_ip_i\|_1\le \epsilon$.

Let $p\in P^\pm(a)$, so that $a^+=J_p(a)$, and let $a_n\in \overline{\spn}(P(a))$ be such that $a_n\to a$ in norm,
 then also $J_p(a_n)\in \overline{\spn}(P(a))$ and $J_p(a_n)\to J_p(a)=a^+$. The proof for $a^-$ and $|a|$ is similar.

\end{proof}

\begin{corollary}\label{coro:comparability} Assume that $(J_p)_{p\in P}$ has the comparability property. 
\begin{enumerate}
\item[(i)] For any $a\in A$ and $p\in P$, $a\in C(p)$ if and only if $P(a)\subseteq C(p)$.
\item[(ii)] For any $a\in A$, there is some block $B\subseteq P$ such that $a\in C(B)$.
\item[(iii)] For any block $B\subseteq P$, $C(B)=\overline{\spn}(B)$.

\end{enumerate}

\end{corollary}

\begin{proof} For (i), let $a\in A$ and $p\in P$. It is clear that if $a\in C(p)$ then $P(a)\subseteq C(p)$. Conversely,
assume that $P(a)\subseteq C(p)$, then $q=J_p(q)+J_{1-p}(q)$ for all $q\in P(a)$, hence for all elements of $\overline{\spn}(P(a))$, the
statement (i) now follows from Theorem \ref{thm:span}. 

Since $P(a)$ is a boolean subalgebra in $P$, it is contained in some block $B\subseteq P$. Then $P(a)\subseteq C(B)$ and
 by Lemma \ref{lemma:CQ_subspace} also $\overline{\spn}(P(a))\subseteq C(B)$, this shows (ii).

For (iii) let $a\in C(B)$, then $B\subseteq PC(a)$, so that $P(a)\subseteq PC(B)=B$, the last equality follows by
 maximality of $B$. We obtain 
\[
a\in \overline{\spn}(P(a))\subseteq \overline{\spn}(B),
\] 
so that $C(B)\subseteq  \overline{\spn}(B)$. The converse inclusion is obvious.

\end{proof}

We will now look at the structure of the C-blocks of $A$ in the comparability case.

\begin{lemma}\label{lemma:compar_cblocks} Assume that $(J_p)_{p\in P}$ has the comparability property and 
let $B$ be a block of $P$. Then $C(B)$ is an order unit space and $(\tilde J_p)_{p\in B}$ where $\tilde J_p=J_p|_{C(B)}$ is a compression base in $C(B)$ with the comparability property.

\end{lemma}

\begin{proof}  By Lemma \ref{lemma:CQ_subspace}, $C(B)$ is an order unit space and $(\tilde J_p)_{p\in B}$ is a compression base in $C(B)$.
By Corollary \ref{coro:comparability} (i), we see that $a\in C(B)$ if and only if
$P(a)\subseteq
C(B)$, where $P(a)$ is computed with respect to all of $P$. By maximality, this means that $P(a)\subseteq B$. 
On the other hand, since all elements in $C(B)$ are compatible with all projections in $B$, the set $P_B(a)$, computed
in $C(B)$ with respect to $B$, is all of $B$. Hence $\emptyset\neq P^\pm(a)\subseteq B$ has the required properties also with respect to
$B$. It follows that $(\tilde J_p)_{p\in B}$ has the  comparability property.

\end{proof}

\begin{theorem}\label{thm:coparability_blocks} Let $\{J_p\}_{p\in P}$ be a compression base in $A$ with the comparability
property. Let $B\subseteq P$ be a block. Then there is a  totally
disconnected compact Hausdorff space $X$ such that
\begin{enumerate}
\item[(i)] $B$ is isomorphic (as a Boolean algebra) to the Boolean algebra $\mathcal P(X)$ of all clopen subsets in $X$.
\item[(ii)] $C(B)$ is isomorphic (as an order unit space) to a norm-dense  order unit subspace in  $C(X,\mathbb R)$.
\item[(iii)] If $A$ is norm-complete, then $C(B)\simeq C(X,\mathbb R)$ (as an order unit space).

\end{enumerate}
\end{theorem}

\begin{proof} By Corollary \ref{coro:comparability}, $C(B)$ is the norm-closed linear span
of $B$. Let $X$ be the Stone space of the block $B$ of $P$, so that $X$ is a totally disconnected compact
Hausdorff space such that (i) is satisfied. Let $\phi: B\to \mathcal P(X)$ be the boolean algebra isomorphism.
Since $B$ is a Boolean algebra, any element  $a\in \spn(B)$ can be (uniquely) written as $a=\sum_{i=1}^k a_ip_i$, 
 where $p_i\in B$ and $\sum_i p_i=1$. In this case, $a\in A^+$ iff $a_i\ge 0$ for all $i$ and $\|a\|_1=\max_i |a_i|$.
Similarly, let $F(X,\mathbb R)$ be the set of all
simple functions in $C(X,\mathbb R)$. Since the clopen sets separate points of $X$, $F(X,\mathbb R)$ is norm-dense in
$C(X,\mathbb R)$ by Stone-Weierstrass theorem.  Any $f\in F(X,\mathbb R)$ can be written as
 $f=\sum_i a_i \chi_{X_i}$, where $\{X_i\}_{i=1}^k$ is a decomposition of $X$ and $\chi_{X_i}$ is the characteristic
function of $X_i$.  So the map $\phi$ extends to a
bijection $\phi: \spn(B)\to F(X,\mathbb R)$, given as
\[
\phi(\sum_i a_i p_i)=\sum_i a_i \chi_{\phi(p_i)}.
\] 
Let $a=\sum_i a_i p_i$, $b=\sum_i b_i q_i$, where $p_i,q_i\in B$, $\sum_i p_i=\sum_j q_j=1$. Then 
$p_i=\vee_j (p_i\wedge q_j)=\sum_j p_i\wedge q_j$, $q_j=\sum_i p_i\wedge q_j$ and we have 
\[
a+b=\sum_{i,j}(a_i+b_j)p_i\wedge q_j, \qquad p_i\wedge q_j\in B,\quad \sum_{i,j} p_i\wedge q_j=1.
\]
Hence
\[
\phi(a+b)=\sum_{i,j} (a_i+b_j)\chi_{\phi(p_i\wedge q_j)}=\sum_{i,j}(a_i+b_j)\chi_{\phi(p_i)}\chi_{\phi(q_j)}=
\phi(a)+\phi(b),
\]
so that $\phi$ is a linear isomorphism of $\spn(B)$ onto $F(X,\mathbb R)$. It is also clear that $\phi$ is a unital  order and
norm isomorphism. Extending $\phi$ to $\overline{\spn}(B)$, we obtain (ii), (iii) is clear.

\end{proof}

In the comparability case, we can extend the notion of compatibility to all pairs of elements in $A$.

\begin{definition}\label{def:compar_compatibility} Let $(J_p)_{p\in P}$ be a compression base in $A$ with the comparability
property. We say that $a,b \in A$ are \emph{compatible} if  $p$ and $q$ are compatible for all $p\in P(a)$ and $q\in
P(b)$.

\end{definition}

Note that by Corollary \ref{coro:comparability} (i), $a$ and $b$ are compatible iff $a\in C(P(b))$ or, equivalently,
 $b\in C(P(a))$. This also shows that if $a$ or $b$ is in $P$, we obtain the same notion of compatibility as in
Definition \ref{def:compatibility}.

\begin{prop}\label{prop:compar_cblocks} Let $(J_p)_{p\in P}$ be a compression base in $A$ with the comparability
property. Then:
\begin{enumerate}
\item[(i)] Two elements in $A$ are compatible if and only if they are in the same C-block.

\item[(ii)] The C-blocks of $A$ are precisely the maximal sets of mutually compatible elements.
\end{enumerate}

\end{prop}

\begin{proof} By Corollary \ref{coro:comparability} (i) and the definition, we see that $a$ and $b$ are compatible iff 
 $P(a)$ and $P(b)$ are contained in the same block $B$, which is equivalent to $a,b\in C(B)$. This proves (i).
To prove (ii), note that by (i), the elements on any C-block are mutually compatible. On the other hand, if 
 $c\in A$ is compatible with all elements of $C(B)$, then clearly $c\in C(B)$.  

\end{proof}

\begin{example}\label{ex:vNcblocks} Let $A=\mathcal M^{sa}$ be the self-adjoint part of a von Neumann algebra $\mathcal
M$. For any projection $p\in \mathcal M$, let $U_p:A\to A$ be defined as $U_p(a)=pap$, $a\in A$. Then $(U_p)_{p\in
\mathcal P}$ is a compression base, where $\mathcal P$ is the set of all projections in $\mathcal M$. It is easily seen
that this compression base has the comparability property.

The elements $a,b\in A$ are compatible in the sense of Definition \ref{def:compar_compatibility} if and only if they
commute.  Blocks of $\mathcal P$ are the maximal sets of
mutually commuting projections in $\mathcal M$ and the C-blocks are precisely the maximal abelian von Neumann
subalgebras in $\mathcal M$.

\end{example}

\subsubsection{Spectrality}

\begin{definition}\label{de:spectral} {\rm \cite[Def. 1.7]{FPspectres}}  The compression base $(J_p)_{p\in P}$ in an order unit space  is \emph{spectral} if it has both the projection cover and the comparability property.
A \emph{spectral order unit space} is an order unit space with a spectral compression base.
\end{definition}

Recall that we say that $P$ is monotone $\sigma$-complete if
suprema (infima) of ascending (descending) sequences in $P$ exist and belong to $P$.

\begin{theorem}\label{thm:spectr_comp} Assume that $P$ is monotone $\sigma$-complete. Then $(J_p)_{p\in P}$ is spectral if
and only if it has the comparability property.

\end{theorem}

\begin{proof} Assume that $P$ is monotone $\sigma$-complete and let $(J_p)_{p\in P}$ have the comparability property.
We will show that $(J_p)_{p\in P}$ also has the projection cover property. So let $a\in E$. By considering the sequence 
 $\lambda_n=1-1/n$ and $1-a$ in Lemma \ref{lemma:descending}, we obtain a descending sequence $q_n\ge q_{n+1}$
 such that $q_n\in P^\pm(1/n-a)$. Then $J_{q_n}(1/n-a)\ge 0$, so that $J_{q_n}(a)\le 1/n q_n$. Put $q=\wedge_n q_n$,
then
\[
J_q(a)=J_qJ_{q_n}(a)\le 1/n J_q(q_n)=1/n q,\qquad \forall n\in \mathbb N,
\]
it follows that $J_q(a)= 0$ and $a\le 1-q$. Assume that $p\in P$ is such that $a\le p$, then $a\in C(p)$ so that 
 by Corollary \ref{coro:comparability} (i), $q_n\in C(p)$ for all $n$. Since $J_{q_n'}(a)\ge 1/nq_n'$, we obtain
\[
1/n J_{p'}(q_n')\le J_{p'}J_{q_n'}(a)=J_{q_n'}J_{p'}(a)=0,
\]
hence $J_{p'}(q_n')=0$, so that $q_n'\le p$ for all $n$.   It follows that $1-q=\vee_n q_n'\le p$, so
that $1-q$ is the projection cover of $a$. This shows that $(J_p)_{p\in P}$ is spectral.

\end{proof}

\begin{lemma}\label{lemma:cxspectral} Let $X$ be a compact Hausdorff space. The following are equivalent.
\begin{enumerate}
\item[(i)] $C(X,\mathbb R)$ is spectral.
\item[(ii)] $C(X,\mathbb R)$ is monotone $\sigma$-complete.
\item[(iii)] $X$ is basically disconnected.
\item[(iv)] The Boolean algebra $\mathcal P(X)$ is monotone $\sigma$-complete.

\end{enumerate}

\end{lemma}

\begin{proof} The equivalences (i) - (iii) were proved in \cite[Thm. 4.8]{FPmonot}. The equivalence (iii) $\iff$ (iv) is well
known.

\end{proof}


\begin{example}\label{ex:finitevalued} (cf. \cite[Ex. 1.7]{FPspectres}) Let $X$ be a totally disconnected compact
Hausdorff space which is  not basically disconnected and let $A= F(X,\mathbb R)$, with the compression base 
 $(J_p)_{p\in \mathcal P(X)}$, where $J_p=U_p|_A$, see Example \ref{ex:CXR}. Then $\mathcal P(X)$ is not monotone
$\sigma$-complete, but it is easily checked that the compression base is spectral.

\end{example}

As we will see next, the property that $A$ is not norm complete was crucial in the above example.

\begin{theorem}\label{thm:cblocks} Assume that $A$ is norm complete and let $(J_p)_{p\in P}$ be a compression base with
the comparability property. The following are equivalent.
\begin{enumerate}
\item[(i)] $(J_p)_{p\in P}$ is spectral.
\item[(ii)] For any block $B$ of $P$, $(\tilde J_p=J_p|_{C(B)})_{p\in B}$ is a spectral compression base in $C(B)$.
\item[(iii)] Any C-block in $A$ is monotone $\sigma$-complete.
\item[(iv)] Any C-block in $A$ is isomorphic to $C(X,\mathbb R)$ for some basically disconnected compact Hausdorff space
$X$.
\item[(iv)] $P$ is monotone $\sigma$-complete.

\end{enumerate}

\end{theorem}

\begin{proof} Assume that (i) holds and let $B$ be a block of $P$, then by Corollary \ref{coro:comparability} (i), $P(a)\subseteq B$ if and only if
$a\in C(B)$. By Lemma \ref{lemma:pc}, $a^0\in P(a)\subseteq B$, so that $(\tilde J_p)_{p\in B}$ has the projection cover
property. We have the comparability property by
Lemma \ref{lemma:compar_cblocks}. This proves (ii).

Assume (ii). By Theorem \ref{thm:coparability_blocks} (iii), any C-block $C(B)$ is isomorphic to $C(X,\mathbb R)$, where $X$ is the Stone space of 
$B$. It is clear that in  this isomorphism, $(\tilde J_p)_{p\in B}$ is identified with the compression base $(U_p)_{p\in
\mathcal P(X)}$, see Example \ref{ex:CXR}.  The equivalence of (ii), (iii) and (iv) now follows by Lemma
\ref{lemma:cxspectral}.

Assume (iii),  we will prove that then every descending sequence in $P$ has an infimum in $P$. We use a method similar
to the one used in \cite[Lemma 1.3]{wright}. 
So let $(p_n)$ be such a sequence and let $L\subseteq P$ be the set of all lower bounds of $(p_n)$ in $P$. Notice that
$L$ is upward directed. Indeed, let $p,q\in L$ and let $a=\frac12(p+q)$. Then $a$ is compatible with each $p_n$, so that 
 there is a block $B$ in $P$ such that $(p_n)\cup \{a\}\subseteq C(B)$. 
Since $C(B)$ is spectral, $a$ has a projection cover $r=a^0\in B$. Since $a\le p_n$, we have $r\le p_n$ for all $n$, so
that $r\in L$. Also $p,q\le 2a\le 2r$, we have $p,q\le r$ by the facial property of $r$. Hence $L$ is upward directed.

Further, let $L_0$ be an increasing chain in $L$, then again $L_0\cup (p_n)$ is contained in some block $B_0$ of $P$.
Since $B_0$ is monotone $\sigma$-complete, $(p_n)$ has an infimum $p_0$ in $B_0$. Clearly, $p_0\in L$ and 
 $p_0\ge q$ for any $q\in L_0$. It follows that every increasing chain in $L$  has an upper bound, so by Zorn's Lemma,
 $L$ has a maximal element, but since $L$ is upward directed, $L$ has a greatest element, which is the greatest lower
bound for $(p_n)$.  This proves (iv).

The implication (iv) $\implies$ (i) is proved in Theorem \ref{thm:spectr_comp}.

\end{proof}

In the rest of this section we assume that $(J_p)_{p\in P}$ is a spectral compression base in $A$.
In such a case,  there is a unique mapping $*: A\to P$, called the \emph{Rickart
mapping}, such that for all $a\in A$ and $p\in P$
\[
p\leq a^*\,\Leftrightarrow\, a\in C(p) \text{ and } J_p(a)=0.
\]
 Indeed, put
\[
a^*:=1-(\|a\|_1^{-1}|a|)^0,\qquad a\in A.
\]
If $p\le a^*$, then $(\|a\|_1^{-1}|a|)^0\le 1-p$. It follows that $a^+,a^-\le \|a\|_1(1-p)$ and hence $J_p(a)=0$, $J_{1-p}(a)=a$, so that $a\in
C(p)$. For the converse, assume that $a\in C(p)$ and $J_p(a)=0$. Let $q\in P^\pm(a)$, then $q\in C(p)$ and
\[
0=J_p(a)=J_{p\wedge q}(a)+ J_{p\wedge q'}(a),
\]
since $0\le J_{p\wedge q}(a)\le q$ and $0\le -J_{p\wedge q'}(a)\le 1-q$, we must have $J_{p\wedge q}(a)=J_{p\wedge
q'}(a)=0$. Hence $J_p(a^+)=J_p(a^-)=J_p(|a|)=0$, so that $|a|\in \Ker^+J_p$. By \eqref{eq:kerj} and the facial property,
we see that $|a|\le \|a\|_1(1-p)$, which entails that $\|a\|_1^{-1}|a|\le 1-p$. This implies that $p\le
1-(\|a\|_1^{-1}|a|)^0=a^*$. Uniqueness is clear.

We note that if an order unit space has the comparability property, then the projection cover property is equivalent to the existence of the Rickart mapping
\cite[Thm 2.1]{FPspectres}\cite[Thm. 6.5]{Fgc}.

Some important properties of the Rickart mapping are collected in the following lemma.

\begin{lemma}\label{le:rickmap} {\rm \cite[Lemma 2.1]{FPspectres}} 
\begin{enumerate}
\item[(i)] For $0\le a\le b$, we have $b^*\le a^*$;
\item[(ii)] For $a\in A$, $a^*\in P(a)$;
\item[(iii)] For $a\in A$, $a^{**}:=(a^*)^*=1-a^*$ is determined by the condition
\[
J_p(a)=a \,\Leftrightarrow \, a^{**}\leq p;
\]
\item[(iv)] For $e\in E$, $e^{**}$ is the projection cover of $e$.

\end{enumerate}

\end{lemma}

\begin{prop}\label{prop:rickart} For $a\in A$, $(a^+)^{**}$ is the least element in $P^\pm(a)$.

\end{prop}

\begin{proof} Let $p=(a^+)^{**}$ and $q\in P^\pm(a)$, so that 
 $a^+=J_q(a)$, $a^-=-J_{1-q}(a)$. Then $J_q(a^+)=a^+$, so that $p\le q$. Hence $1-q\le 1-p$, so that 
$J_p(a^-)=0$. It follows that $J_p(a)=J_p(a^+)=a^+\ge 0$ and $J_{1-p}(a)=J_{1-p}(-a^-)=-a^-\le 0$. 
In particular, $p\in PC(a)$. We next show that $p\in PC(PC(a))$. 

Assume that $q\in PC(a)$ and let $r\in P^\pm(a)$. Then $r$ is compatible with $q$ and $r,q\in PC(a)$, so that by
 Lemma \ref{lemma:pca}, we obtain
\[
a=(J_{q\wedge r}+J_{q'\wedge r}+J_{q\wedge r'}+J_{q'\wedge r'})(a),
\]
so that
\[
a^+=J_r(a)= J_{q\wedge r}(a)+J_{q'\wedge r}(a)= J_q(a^+)+J_{q'}(a^+).
\]
It follows that $PC(a)\subseteq PC(a^+)$. Using Lemma \ref{le:rickmap}, we obtain
\[
p\in P(a^+)\subseteq PC(PC(a^+))\subseteq PC(PC(a)).
\]
Now we see that  $p\in P^\pm(a)$ and we have already shown that $p\le q$ for all $q\in P^\pm(a)$.   

\end{proof}

Similarly as in the case of spectral duality, we may now define the  \emph{spectral resolution} of $a$ as a family
 $(p_{a,\lambda})_{\lambda \in {\mathbb R}}\subseteq P$ defined by \cite[Def. 3.2]{FPspectres}
\[
p_{a,\lambda}:=((a-\lambda)^+)^*.
\]
Note that by Lemma \ref{le:rickmap}, $p_{a,\lambda}\in P(a)$ for all $\lambda\in \mathbb R$.

Let $\lambda_1\le\dots \le \lambda_n$ be any sequence. Using Proposition \ref{prop:rickart} and  the
construction in the proof of Lemma \ref{lemma:descending}, we see that $q_i:=((a-\lambda_i)^+)^{**}$ is a descending
sequence in $P$, which by Lemma \ref{le:rickmap} (iii) gives an  ascending sequence of spectral projections
\[
p_{a,\lambda_1}\le\dots \le p_{a,\lambda_n}.
\] 
Proceeding as in the proof of Theorem \ref{thm:span}, we can see that every $a\in A$ can be written as a norm-convergent
integral  of Riemann-Stieltjes type
\[
a=\int_{L_a-0}^{U_a}\lambda dp(\lambda),
\]
where  the \emph{spectral lower and upper bounds} for $a$ are defined by $L_a:=\sup\{ \lambda \in {\mathbb R}:\lambda \leq a\}$ and $U_a:=\inf\{\lambda \in {\mathbb R}:a\leq \lambda \}$, respectively.

Properties of spectral resolutions are described in \cite[Section 3]{FPspectres}. In particular, we have

\begin{corollary}\label{co:commut} If $p\in P$, then $a\in C(p)\,\Leftrightarrow\, p_{a,\lambda}\in C(p)$ for all $\lambda \in {\mathbb R}$.
\end{corollary}

\begin{corollary}\label{co:simple} There exists an ascending sequence $a_1\leq a_2\leq \cdots $ in $C(PC(a))$ such that
each $a_n$ is a finite linear combination of projections in the family $(p_{a,\lambda})_{\lambda \in {\mathbb R}}$ and
$\|a-a_n\|_1 \to 0$.
\end{corollary}

\section{A comparison of the two spectrality theories }\label{sec:comparison}

In this section we compare the spectral theory in order unit spaces presented in \cite{FPspectres} with the theory of Alfsen and Shultz \cite{AlSh}.
We show in more details that the Alfsen and Shultz  theory may be considered as a special case of the theory presented in \cite{FPspectres}.

%
%
%
%

\subsection{Compressions and F-compressions}

We start by describing more properties of F-compressions that can be obtained taking the $\sigma(A,V)$-continuity into account.

Let $J:A\to A$ be an F-compression with focus $p$ and let $J^*:V\to V$ be its adjoint:
\[
J^*:V\to V,\quad J^*\varphi=\varphi\circ J,\qquad \varphi\in V.
\]
It is easy to see that $J^*$ is $\sigma(A,V)$-continuous, positive and idempotent. Moreover, we have
\begin{equation}\label{eq:duals}
\Ker^+J^*=\{p\}^\bullet,\qquad \Iim^+J^*\subseteq \{1-p\}^\bullet=(\Iim^+J^*)^{\bullet\bullet}.
\end{equation}
Indeed, we have
\begin{align*}
\Ker^+ J^*&=\{\varphi\in V^+:\ J^*\varphi=0\}=\{\varphi\in V^+:\ \<J(a),\varphi\>=0,\ \forall a\in A^+\} \\
&=(\Iim^+J)^\bullet=\{p\}^\bullet
\end{align*}
and similarly
\begin{align*}
\Ker^+ J&= \{a\in A^+:\ J(a)=0\}=\{a\in A^+:\ \<J(a),\varphi\>=0,\ \forall \varphi\in V^+\} \\
&=\{a\in A^+:\ \<a,J^*\varphi\>=0,\forall \varphi\in V^+\}=(\Iim^+J^*)^\bullet,
\end{align*}
so that
\[
\Iim^+J^*\subseteq (\Iim^+J^*)^{\bullet\bullet}=(\Ker^+J)^\bullet=\{1-p\}^\bullet.
\]

\begin{lemma}\label{lemma:semiexp} Let $p$ be the focus of an F-compression. Then 
\[
\{1-p\}^{\bullet\bullet}\cap E= [0,1-p].
\]

\end{lemma}

\begin{proof} As we have seen, 
\[
\face_{A^+}(1-p)=\Ker^+J=(\Iim^+J^*)^\bullet=\{1-p\}^{\bullet\bullet}.
\] 
The result follows by Lemma \ref{lemma:principal}.

\end{proof}


For an F-compression $J$ with focus $p$, we denote
\begin{align*}
K_p&:=\{1-p\}^\bullet\cap K=\{\rho\in K,\ \<p,\rho\>=1\},\\
\mathcal S(J)& := \Iim(J^*)\cap K.
\end{align*}

Then $\mathcal S(J)$ is a base of the subcone $\Iim^+J^*\subseteq V^+$ and
by \eqref{eq:duals}, 
\begin{equation}\label{eq:duals_base}
\mathcal S(J)\subseteq K_p= \mathcal S(J)^{\bullet\bullet}\cap K.
\end{equation}

\begin{lemma}\label{le:comp_determined} Let $J$ be an F-compression. Then $J$ is uniquely determined by its focus $p$
 and the subset $\mathcal S(J)$.

\end{lemma}

\begin{proof} Any idempotent map  $J$ is uniquely determined by its image and kernel. Since $\Iim(J)$ is positively generated
 and $\Iim^+J=\face_{A^+}(p)$, we see that
\[
\Iim(J)=\{a\in A:\ \exists \lambda\ge 0,\ -\lambda p\le a\le \lambda p\}
\]
is the order ideal in $A$ generated by $p$. Further, it is easy to see that 
\[
\Ker(J)=\Iim(J^*)^\circ=(\Iim^+J^*)^\circ=\mathcal S(J)^\circ,
\]
since $\Iim(J^*)$ is positively generated as well and $\mathcal S(J)$ is a base of $\Iim^+J^*$.

\end{proof}

Following \cite{AlSh}, we introduce an important property. A continuous positive  idempotent map    $J:A\to A$ is 
\emph{smooth} if
\[
\Ker(J)=\Tan_{A^+}(\Ker^+J),
\]
where $\Tan_{A^+}(\Ker^+J)$ is the \emph{tangent space} of $A^+$ at $\Ker^+J$. 
If $J$ is an F-compression with focus $p$, then we have (see \cite[Prop. 7.3]{AlSh})
\[
 \Tan_{A^+}(\Ker^+J)=\Tan_{A^+}(1-p)=\{1-p\}^{\bullet\circ}.
\]
We say that the dual map $J^*:V\to V$ is \emph{neutral} if
\[
\rho\in K_p \implies \<J(a),\rho\>=\<a,\rho\>,\ \forall a\in A.
\]
\begin{lemma}\label{lemma:smooth} Let $J$ be an F-compression with focus $p$.
The following are equivalent.
\begin{enumerate}
\item[(i)] $J$ is smooth;
\item[(ii)] $\mathcal S(J)=K_p$;
\item[(iii)] $\Iim^+J^*=\{1-p\}^\bullet$;
\item[(iv)] $\Iim^+J^*$ is a semi-exposed face of $V^+$;
\item[(v)] $J^*$ is neutral.

\end{enumerate}
\end{lemma}

\begin{proof}
It is clear that (ii) is equivalent to (iii), the equivalence of (iii) and (iv)
 is clear from \eqref{eq:duals}. The fact that (i) is equivalent to (iv) and (v) was proved in \cite[Chap. 7]{AlSh} for more general maps, 
we give an easy proof for F-compressions. Note that (v) can be reformulated as:
\[
 \varphi\in V^+,\ \<1-p,\varphi\>=0 \implies J^*\varphi=\varphi,
\]
in other words, $\{1-p\}^\bullet\subseteq \Iim^+J^*$. Again by \eqref{eq:duals}, this is equivalent to (iii). 
Observe that
\begin{equation}\label{eq:smooth}
\{1-p\}^{\bullet\circ}\subseteq (\Iim^+J^*)^\circ=\{a\in A:\ \<J(a),\varphi\>=0,\ \forall \varphi\in V^+\}=\Ker(J)
\end{equation}
and (i) means that equality holds in \eqref{eq:smooth}. It is now clear (iii) implies (i). Conversely, assume (i), then 
\[
\{1-p\}^{\bullet}\subseteq \{1-p\}^{\bullet\circ\circ}\cap V^+=(\Iim^+J^*)^{\circ\circ}\cap V^+=\Iim(J^*)\cap V^+= \Iim^+ J^*,
\]
here we used the fact that $\Iim(J^*)$ is a closed linear subspace generated by $\Iim^+
J^*$, so that $(\Iim^+J^*)^{\circ\circ}= \Iim(J^*)$. The proof now follows by \eqref{eq:duals}.

\end{proof}

\begin{lemma}\label{lemma:smooth_unq} 
Let $p\in E$ be the focus of a smooth F-compression $J$. Then $J$ is the unique F-compression with focus $p$.

\end{lemma}

\begin{proof} Let $\tilde J$ be an  F-compression with focus $p$. Then $\Iim(\tilde J)=\Iim(J)$ and 
\[
\Ker(J)=K_p^\circ\subseteq \mathcal S(\tilde J)^\circ=\Ker(\tilde J).
\]
But then for any $a\in  A$,
\[
\tilde J(a)=\tilde J(J(a)+a-J(a))=\tilde J(J(a))=J(a).
\]

\end{proof}

We can now compare  compressions (Definition \ref{de:compr}) and F-compressions (Definition \ref{de:Fcompr}).

\begin{theorem}\label{thm:AS_FP} Any compression is an F-compression. An F-compression is a compression if and only if
it is smooth and has a smooth complementary F-compression.

\end{theorem}

\begin{proof}  Let $J:A\to A$ be a compression. Since $J$ is a normalized 
 idempotent $\sigma(A,V)$-continuous linear mapping, we only need to check the properties (F2) and (F3) of Definition \ref{de:Fcompr}. 

Let $J'$ be the complement of $J$, then  $J'(1)=1-p$, this can be seen similarly as in Lemma \ref{le:complementary}.
 To show (F2), let $0\leq e\leq p$, then $J'(e)\leq J'(p)=0$, so $e\in
\Ker^+ J'=\Iim^+J$, whence $J(e)=e$. For (F3), observe that  $J(e)=0$ implies $e\in \Ker^+ J=\Iim^+J'$. Hence 
$J'(e)=e\leq 1$ so that  $e=J'(e)\leq J'(1)=1-p$.


By Definition \ref{de:compr}, a continuous positive idempotent map  $J$ is a compression if and only if it is bicomplemented, 
which means that there is a
complementary continuous positive idempotent map $J'$ such that the dual maps $J^*$ and $(J')^*$ are also complementary.
In this case, $J'$ is a compression as well. In particular, $J$ and $J'$ are complementary F-compressions such that 
 by \eqref{eq:duals}
\[
\{p\}^\bullet=\Ker^+J^*= \Iim^+(J')^*,\qquad \{1-p\}^\bullet =\Ker^+(J')^*= \Iim^+J^*.
\] 
By Lemma \ref{lemma:smooth}, this means that both $J$ and $J'$ are smooth (cf. \cite[Prop. 7.20]{AlSh}).

\end{proof}

%
%
\subsection{Compression bases and spectrality}

Recall that the focus of a compression is called a projective unit. 
Let $\mathcal P$ denote  the set of all projective units in $A$ and 
for each $p\in \mathcal P$, let $J_p$ be the unique compression with focus $p$. The set $\mathcal J=(J_p)_{p\in \mathcal P}$ 
of compressions has a natural ordering given as $J_p\preceq J_q$ if $J_qJ_p=J_p$ and the map $J_p\mapsto p$ is an order
isomorphism onto the set $\mathcal P$ with the ordering induced from $A$, \cite[Prop. 7.32]{AlSh}.

In general, it is not clear whether $(J_p)_{p\in \mathcal P}$ is a compression base in $A$. Therefore, although all
$J_p$ are F-compressions, we cannot directly use the properties of the map $p\mapsto J_p$ proved for compression bases. 
We will next show that it is the case if the Standing Hypothesis \ref{ass:SH} is true, recall that this means
that  for any $a\in A^+$ there is a 
 projective unit denoted by $r(a)$ such that the exposed face of $K$ (or, equivalently, of  $V^+$) determined by $a$ is
also determined by $r(a)$, that is 
\[
\{a\}^{\bullet}=\{r(a)\}^\bullet.
\]
It is proved in  \cite[Prop. 8.1]{AlSh} that under this assumption,  $\mathcal P$ is an orthomodular
 lattice and by the above order isomorphism, 
 $\mathcal J$ (with complementation) is an OML as well,  with $J_p\vee J_q=J_{p\vee q}$ and $J_{p\wedge q}=J_p\wedge J_q$. 
Moreover,   two compressions $J_p$ and $J_q$ commute if and only if $J_p(q)\le q$ and  then 
$J_pJ_q=J_p\wedge J_q=J_{p\wedge q}$, \cite[Thm. 8.3]{AlSh}.

%

\begin{theorem}\label{th:normal} If the Standing Hypothesis is satisfied, then $\{J_p\}_{p\in \mathcal P}$ is 
a compression base in $A$, with the projection cover property.

\end{theorem}

\begin{proof} We first prove that $\mathcal P$ is a normal subalgebra in $E$. If $p+q\in E$, then $p\le 1-q$ and
 $p+q=p\vee q=(J_p\vee J_q)(1)\in \mathcal P$, so that $\mathcal P$ is a subalgebra. Further, let
 $e,f,d\in E$, $e+d+f\leq 1$, $p:=e+d, q:=f+d\in {\mathcal P}$.  Then $f\leq 1-p,\ d\leq p$ implies that
$J_p(q)=J_p(f)+J_p(d)=d\le q$ so that $J_p$ and $J_q$ commute and  we obtain 
\[
d=J_pJ_q(1)=J_q\wedge J_p(1)\in \mathcal P.
\]
Similarly, if $p,q,r\in \mathcal P$  and $p+q+r\le 1$, then 
$J_{p+r}J_{q+r}=J_{q+r}J_{p+r}=J_r$. Hence $(J_p)_{p\in \mathcal P}$ is a compression base. Let now $a\in E$ and let
$p=r(a)$. This implies that 
\[
a\in \{a\}^{\bullet\bullet}\cap E=\{p\}^{\bullet\bullet}\cap E=[0,p],
\]
hence $a\le p$, here we have used Lemma \ref{lemma:semiexp}.  If $a\le q\in \mathcal P$,
 then clearly $\{q\}^\bullet\subseteq \{a\}^\bullet=\{p\}^\bullet$, so that 
\[
p\in  \{p\}^{\bullet\bullet}\cap E\subseteq \{q\}^{\bullet\bullet}\cap E=[0,q],
\]
hence $p\le q$. It follows that $p=a^0$. (Also \cite[Prop. 8.24]{AlSh}: $p$ is the least projective unit such that 
$a\in \face_{A^+}(p)$).

\end{proof}

\begin{theorem}\label{th:gencomp} Assume that  $A=V^*$. Then the following are equivalent:
\begin{enumerate}
\item[(i)] $A$ and $V$ are in spectral duality;
\item[(ii)] $(J_p)_{p\in \mathcal P}$ is a spectral compression base;
\item[(iii)] There is a spectral compression base $(J_p)_{p\in P}$ such that all $J_p$ are smooth.

\end{enumerate}

\end{theorem}

\begin{proof} By \cite[Thm. 8.52,Thm. 8.55]{AlSh}, $A$ and $V$ are in spectral duality if and only if the Standing Hypothesis holds and every
element $a\in A$ has a unique orthogonal decomposition $a=a^+-a^-$. So assume (i) spectral duality. We already know by 
Theorem \ref{th:normal} that $(J_p)_{p\in \mathcal P}$ is a compression base with the projection cover property. 
To  show comparability, let $a\in A$ and let $a=a^+-a^-$ be the unique orthogonal decomposition. By \cite[Thm.
8.57]{AlSh}, $J_r$ with  $r=r(a^+)$ is the least compression such that $J_r(a)\ge 0$ and $J_r(a)\ge a$, which implies
that $J_{1-r}(a)\le 0$. Moreover, by \cite[Thm. 8.62]{AlSh}, $r\in C(PC(a))$. Hence $r\in P^\pm(a)$ and
$(J_p)_{p\in \mathcal P}$ has the comparability property, so (ii) is true. 

The implication (ii) $\implies$ (iii) follows by Theorem \ref{thm:AS_FP}, so assume (iii). Then by Theorem \ref{thm:AS_FP}, all $J_p$ are
compressions and so $\mathcal P\supseteq P$. Moreover, by Lemma \ref{le:principal}, all sharp elements are in $P$, so that 
 $\mathcal P=P$ and $(J_p)_{p\in P}$ is the set of all compressions.
Let $a\in A$. By Proposition \ref{prop:rickart}, $r=(a^+)^{**}$ is the least element in $P^\pm(a)$ and
since $p\mapsto J_p$ is an order isomorphism, $J_r$ is the least compression $J$ such that $J'(a)\le 0\le J(a)$,
equivalently, $J(a)\ge a$ and $J(a)\ge 0$. This implies (i) (see Definition \ref{def:spectral}).

\end{proof}

\begin{corollary} Let $A$ and $V$ be in spectral duality. Then F-compressions coincide with compressions.

\end{corollary}

\begin{proof} Let $A$ and $V$ be in spectral duality. We already know that every compression is an F-compression
and that $(J_p)_{p\in \mathcal P}$ is a spectral compression base. By Lemma \ref{le:principal},
every sharp element in $E$ is in $\mathcal P$ and therefore is a projective unit. 
Since the focus of an F-compression is sharp, it is a projective unit. The statement  follows by Lemma \ref{lemma:smooth_unq}.

\end{proof}

\section{Spectrality for JB-algebras }\label{sec:JB}

We describe an important example of (spectral) compression bases on JB-algebras. An overall reference we use 
for JB-algebras  is \cite[Chap. 1]{AlSh}.

\begin{definition}\label {de:jordan} {\rm \cite[Definition 1.1]{AlSh}} A \emph{Jordan algebra }  $A$ over $\mathbb R$ is a vector space equipped with a commutative bilinear product $\circ$ that satisfies the identity
\begin{equation}\label{eq.1}
(a^2\circ b)\circ a =a^2\circ(b\circ a) \ \mbox{for all}  \ a,b\in A.
\end{equation}

\end{definition}

\begin{definition}\label{de:JBlg} \rm{ \cite[Definition 1.5]{AlSh}} A JB-algebra is a Jordan algebra $A$ over $\mathbb
R$ with identity element $1$ equipped with a complete norm satisfying the following requirements for $a,b\in A$:
\begin{eqnarray*}
\|a\circ b\| &\leq& \|a\|\|b\|,\\
\|a^2\|&=& \|a\|^2,\\
\|a^2\|&\leq & \|a^2 +  b^2\|.
\end{eqnarray*}
\end{definition}

\begin{example}\label{ex:1} The self-adjoint part  $A^{sa}$ of a C*-algebra $A$ is a JB-algebra with Jordan product $a\circ b=\frac{1}{2}(ab+ba)$. Any norm closed Jordan subalgebra of ${\mathcal B}(H)^{sa}$ is a JB-algebra.
\end{example}

Let $A$ be a JB-algebra. The positive cone in $A$ is defined as
\[
A^+=A^2=\{a^2:a\in A\}.
\]
Then $(A,A^+,1)$ is a norm-complete  order unit space and its order unit norm $\|\cdot\|_1$ coincides with the norm $\|\cdot\|$ in $A$,
\cite[Theorem 1.11]{AlSh}. An element $p\in A$ is called a \emph{projection} if $p=p^2$. The set of all projections in
$A$ will be denoted by $P$. Clearly, we then have $p\in A^+$ and $(1-p)^2=1-p$, so that $1-p\in P$ and $p\le 1$.

The Jordan triple product in $A$ is defined as
\[
\{ abc\} := (a\circ b)\circ c + (c\circ b)\circ a - (a\circ c)\circ b),\qquad a,b,c\in A.
\]
If $a=c$ we have
\[
\{aba\}=2(a\circ b)\circ a-a^2\circ b.
\]

\begin{lemma}\label{le:1.26} {\rm \cite[Lemmas 1.26 and 1.37]{AlSh}} If $a,b\in A^+$, then 
{\rm(i)}  and {\rm(ii)} are equivalent and imply {\rm(iii)}, where 
\begin{enumerate}
\item[{\rm(i)}] $\{ aba\}=0$,
\item[{\rm(ii)}] $\{ bab\}=0$,
\item[{\rm(iii)}] $a\circ b=0$.
\end{enumerate}
If $a$ is a projection, then all (i) - (iii) are equivalent.
\end{lemma}

For any $a,c\in A$, we define a linear map $U_{a,c}:A\to A$ by
\[
U_{a,c}(b)=\{abc\},\qquad b\in A.
\]
If $a=c$, we put $U_{a}(b):=U_{a,a}(b)$. For any $a\in A$, the map $U_a$ is positive \cite[Theorem 1.25]{AlSh}.
If $p\in P$, we get
\[
U_p(a)=2p\circ(p\circ a)-p\circ a,\qquad a\in A.
\]

\begin{lemma}\label{lemma:JB_ordering} Let $a\in E$, $p\in P$. Then $a\le p$ if and only if $a\circ p=a$ if and only if 
$U_p(a)=a$ if and only if $U_{1-p}(a)=0$.

\end{lemma}

\begin{proof} Using Lemma \ref{le:1.26}, we obtain the following equivalences: 
\[
a\circ p=a\  \iff\ a\circ (1-p)=0\ \iff\ U_{1-p}(a)=0.
\]
On the other hand, it is clear that $a\circ p=a$ implies $U_p(a)=a$ and since $U_p$ is positive and $U_p(1)=p$, we
obtain that $a=U_p(a)\le U_p(1)=p$. From $a\le p$, it similarly follows that $0\le U_{1-p}(a)\le U_{1-p}(p)=0$, which
 implies $a\circ p=a$, closing the chain of implications.

\end{proof}

Elements $a,b\in A^+$ are \emph{orthogonal} ($a\perp b$) if $\{ aba\}=\{bab\}=0$. Note that if $p,q\in P$, then 
$p\perp q$ if and only if $p\le 1-q$ (by Lemma \ref{lemma:JB_ordering}).
 By \cite[Proposition 1.28]{AlSh}, each $a\in A$ can be expressed uniquely as a difference of positive orthogonal elements:
\begin{equation}\label{eq:orthogonal_dec}
a=a^+-a^-,\qquad a^+,a^-\in A^+,\qquad a^+\perp a^-.
\end{equation}
Both $a^+, a^-$ are contained in the norm-closed subalgebra $A(a,1)\subseteq A$ generated by $a$ and $1$. The decomposition in
\eqref{eq:orthogonal_dec} is called the \emph{orthogonal decomposition} of $a$.


From now on, we will consider $A$ in the separating order and norm duality with $V=A^*$.

\begin{prop}\label{prop:JBcompr} For any $p\in P$,  $U_p$ is a compression on $A$ with focus $p$.

\end{prop}

\begin{proof} We will first show that $U_p$ is an F-compression. 
As we have seen, $U_p$ is linear and positive and it is clear that $U_p(1)=p\le 1$, so (F1) holds. 
The conditions (F2) and (F3) follow from Lemma \ref{lemma:JB_ordering}. Let $U_p^*: V\to V$ be the dual map.
By \cite[Prop. 1.41]{AlSh}, $U_p^*$ is neutral, hence $U_p$ is smooth (Lemma \ref{lemma:smooth}). Since $U_{1-p}$ has
the same properties and is complementary to $U_p$ (Lemma \ref{le:complementary}), the proof is finished by Theorem \ref{thm:AS_FP}.

\end{proof}

We have the following characterization of projections.

\begin{lemma}\label{lemma:JB_sharp}
For $p\in E$, the following are equivalent.
\begin{enumerate}
\item[(i)] $p$ is a projection;
\item[(ii)] $p$ is principal;
\item[(iii)] $p$ is  sharp.

\end{enumerate}

\end{lemma}

\begin{proof} Assume that $p\in P$, then by Proposition \ref{prop:JBcompr}, $p$ is the focus of the compression $U_p$.
Hence $p$ is principal, so (i) implies (ii). The implication (ii) $\implies$ (iii) is by Lemma
\ref{lemma:princ_ext_sharp}. 
 Finally, let $p$ be sharp. Since $p\in E$, we have by the functional calculus \cite[Prop. 1.21]{AlSh} that  
 $p^2\le p$. It follows that $0\le p-p^2=(1-p)-(1-p)^2\le p, 1-p$, so that $p-p^2=0$ and $p\in P$.

\end{proof}

It follows by Proposition \ref{prop:JBcompr},  Lemmas \ref{lemma:princ_ext_sharp} and \ref{lemma:smooth_unq} that any 
 F-compression on $A$ is of the form $U_p$ for some projection $p$.  We will next show that also any 
retraction on $A$ is of this form. By \cite{Funig}, this means that the additive group of $A$ is a {\emph compressible
group}, which will show that  $(U_p)_{p\in P}$ is a compression base. We will need some preparation first.

Let $T_a:A\to A$ be the map obtained by Jordan multiplication by $a$. We say that  $a,b\in A$ \emph{operator commute} if
$T_aT_b=T_bT_a$. This will be denoted by $a\leftrightarrow b$. The next result shows that if $b$ or $a$ is a projection,
then this relation  is the same  as compatibility.

\begin{theorem}\label{th::equiv} {\rm \cite[Proposition 1.47]{AlSh}} Let $A$ be a JB-algebra, $a\in A$, and let $p$ be a projection in $A$. The following are equivalent:
\begin{enumerate}
\item[{\rm(i)}] $a$ and $p$ operator commute;
\item[{\rm(ii)}] $T_pa=U_pa$;
\item[{\rm(iii)}] $(U_p+U_{1-p})(a)=a$ (that is, $a\in C(p)$);
\item[{\rm(iv)}] $a$ and $p$ are contained in an associative subalgebra in $A$.
\end{enumerate}
\end{theorem}

It also follows that any element in the subalgebra $A(a,1)$ generated by $a$ and $1$ commutes with any projection that
commutes with $a$, that is, 
\begin{equation}\label{eq:JB_cpc}
A(a,1)\subseteq C(PC(a)).
\end{equation}
Indeed, let $a\in C(p)$, then  $a$ and $p$ are contained in an associative subalgebra $A_0\subseteq A$, which clearly
also contains $A(a,1)$, so that $A(a,1)\subseteq C(p)$.

We will also need the following  \emph{Peirce decomposition} of elements in $A$.

\begin{theorem}\label{th:peirce}{\rm \cite[Thm. 1.4]{Arz}} Let $A$ be a Jordan algebra with unit $1$, $p\in A$ be a projection. Then $A$ decomposes into the direct sum
\[
A=U_p(A)\oplus U_{p,1-p}(A)\oplus U_{1-p}(A),
\]
where
\begin{eqnarray*}
U_p(A)&=&\{x\in A: p\circ x=x\},\\
U_{p,1-p}(A)&=&\{ x\in A: p\circ x=\frac{1}{2}x\},\\
U_{1-p}(A)&=&\{x\in A: p\circ x=0\}.
\end{eqnarray*}
If $x\in U_p(A)$ or $x\in U_{1-p}(A)$ then $x\in C(p)$ (\cite[Prop. 1.5]{Arz}).

\end{theorem}

The proof of the following result is a modification of the proof of \cite[Theorem 4.5]{Fcompog}.

 \begin{theorem}\label{th:compressible}  Let $A$ be a JB-algebra and let $J:A \to A$ be a retraction with focus $p=J(1)$. 
Then $p\in P$ and $J=U_p$.
\end{theorem}

\begin{proof} As the focus of a retraction, $p$ is principal and hence $p\in P$ by Lemma \ref{lemma:JB_sharp}. 
Let $p'=1-p$. If $e\in A$, $0\leq e\leq 1$ then $0\leq \{p'ep'\}\le p'$, whence
\[
0\leq e\leq 1\ \implies \ J(\{p'ep'\})=0.
\]
If $0\leq e\leq 1$ then $0\leq pep\leq p$, whence
\[
0\leq e\leq 1 \ \implies \ J(\{pep\})=\{pep\}.
\]
We claim that for $a\in A$, $J(\{pap\})=\{pap\}=U_p(a)$. Indeed, this is easily seen to be true for all positive elements
 by normalization. Since we can always write 
$a=x-y$ with $x,y\in A^+$, we have $J(\{pap\})=\{pap\}$ for all $A$.  By the same argument, $J(\{p'ap'\})=0$ for $a\in
A$.
For the projection $p$ we have the Peirce decomposition as in Theorem \ref{th:peirce}:
\[
A=U_p(A)\oplus U_{p,1-p}(A)\oplus U_{1-p}(A).
\]
Let $a\in A$, then $a=a_1+a_2+a_3$ where $a_1\in U_p(A)$, $a_2\in U_{p,1-p}(A)$, $a_3\in U_{1-p}(A)$.
The we have
\[
a_1=\{pa_1p\},\quad
 a_2\circ p=\frac{1}{2}a_2,\quad 
a_3=\{p'a_3p'\}.
\]
For $n\in \mathbb Z$, consider
\begin{eqnarray*}
0\leq (a_2-np')^2 &=& a_2^2-2na_2\circ p'+n^2p'\\
&=& a_2^2-na_2+n^2p'.
\end{eqnarray*}
Applying $J$ and taking into account that $J(p')=0$ we get
\[
nJ(a_2)\leq J(a_2^2),\quad  \forall n\in {\mathbb Z}.
\]
For positive $n$ we obtain $J(a_2)\leq 0$, for $n$ negative we obtain $-J(a_2)\leq 0$, hence $J(a_2)=0$.
Consequently,
\[
J(a)=J(a_1)+J(a_2)+J(a_3)=\{pa_1p\}=\{pap\}.
\]

\end{proof}

\begin{corollary}\label{coro:JBcompb} The set $\{U_p\}_{p\in P}$ is a compression base in $A$.

\end{corollary}

\begin{proof} The statement is a special case of \cite[Theorem 2.3]{Funig}. We rewrite the proof in the present
situation for the convenience of the reader.

Assume that $p,q\in P$, $p+q\le 1$, so that $p\le 1-q$. It follows that $p\circ(1-q)=p$ and hence $p\circ q=0$. 
We have $(p+q)^2=p+q+2p\circ q=p+q$, so that $p+q\in P$ and $P$ is a subalgebra in $E$. Let us prove that $P$ is normal.
So let $e,f,d\in E$, $e+f+d\le 1$ and assume that $p:=e+d\in P$ and $q:=f+d\in P$. Put $J:=U_p\circ U_q$, then $J:A\to
A$ is a positive linear map and since $f\le 1-p$ and $d\le p$, we have 
\[
J(1)=U_p(q)=U_p(f+d)=U_p(d)=d\in E.
\]
Further, let $0\le a\le d$, then since $a\le p$ and $a\le q$, we have 
\[
J(a)=U_p(U_q(a))=U_p(a)=a.
\]
Hence $J$ is a retraction and by Theorem \ref{th:compressible}, $d\in P$ and $J=U_d$. This also shows that 
 for $p,q,r\in P$, $p+q+r\le 1$, we have $U_{p+r}\circ U_{q+r}=U_r$.

\end{proof}

In \cite{Arz} the following symbols are defined for a subset $S$ of $A$:
\begin{align*}
S^{\perp}=&\{ a\in A:\ U_a(x)=0,\ \forall x\in S\},\\
^{\perp}S=&\{ x\in A:\  U_a(x)=0,\ \forall a \in S\},\\
^{\perp}S^+=& ^{\perp}S\cap A^+.
\end{align*}

The following notion of a Rickart Jordan algebra was introduced by Ayupov and Arzikulov.

\begin{definition}\label{de:rickart} {\rm \cite[Definition]{AjArz}, \cite[Theorem 1.6, Theorem 1.7]{AjArz}}
 A Jordan algebra $A$ is \emph{Rickart} if
one of the following equivalent statements is true
\begin{enumerate}
\item[(A1)] For every element $x\in A^+$ there is a projection $p\in A$ such that
\[
\{x\}^{\perp}=U_p(A)
\]
\item[(A2)] For every element $x\in A$ there is a projection $p\in A$ such that
\[
^{\perp}\{ x\}^+=U_p(A)^+.
\]
\end{enumerate}

\end{definition}

It was proved in \cite{AjArz} that the self-adjoint part of a C*-algebra $\mathcal A$ is Rickart JB-algebra iff
$\mathcal A$ is a Rickart C*-algebra.

\begin{definition}\label{de:car} An element $p\in P$ is called a \emph{carrier} of  $a\in A$ if it is the smallest projection  such that $p\circ a=a$. The carrier of $a$ will be denoted by $s(a)$.
\end{definition}

The following result is straightforward from Lemma \ref{lemma:JB_ordering}.

\begin{lemma}\label{lemma:JBproj_cover} Let $a\in E$. Then $s(a)$ is the smallest projection such that $a\le s(a)$.

\end{lemma}

\begin{lemma}\label{le:car} {\rm \cite[Lemma 2.7]{Arz}} Let $a\in A^+$, $p\in P$. Then $p=s(a)$ if 
$^{\perp}\{ a\}^+=U_{1-p}(A)^+$.
\end{lemma}
\begin{proof} Note that using Lemma \ref{lemma:JB_ordering}, we have for $q\in P$ and $a\in A^+$ that
 $q\circ a=a$ iff  $(1-q)\circ a=0$ iff $1-q\in {^\perp}\{a\}^+$. It follows that  $1-s(a)$ is the largest
projection in ${^\perp}\{a\}^+$. If $p\in P$ is such that ${^\perp}\{a\}^+=U_{1-p}(A)^+$, then  
  $1-p=U_{1-p}(1-p)\in {^\perp}\{a\}^+$ and for any projection $r\in {^\perp}\{a\}^+$ we have 
 $U_{1-p}(r)=r$, so that $r\le 1-p$ by Lemma \ref{lemma:JB_ordering}. Hence $p=s(a)$.   

\end{proof}

\begin{lemma}\label{le:orthog} Let $A$ be a Rickart JB-algebra. For $a,b\in A^+$ then

{\rm(i)} $\{aba\}=0 \ \Leftrightarrow \ \{s(a)bs(a)\}=0$.

If, in addition, $a\leftrightarrow b$, then

{\rm(ii)} $a\circ b=0 \ \implies \ s(a)\leftrightarrow b$ and $s(a)\circ b=0$.
\end{lemma}

\begin{proof} For (i), note that  $\{aba\}=0$ $\Rightarrow$ $b\in {^{\perp}\{a\}}^+=U_{1-s(a)}(A)^+$, hence $b=U_{1-s(a)}(b)$, which implies $U_{s(a)}(b)=0$, i.e., $\{s(a)bs(a)\}=0$.
If $\{s(a)bs(a)\}=0$, then $\{bs(a)b\}=0$ by Lemma \ref{le:1.26}, and $a\leq \|a\|s(a)$ implies $\{bab\}\leq
\|a\|\{bs(a)b\}=0$ which  yields $\{aba\}=0$. For (ii), note that if $a\leftrightarrow b$, then $a\circ b=0$ implies
$\{aba\}=0$. Indeed, we have
\[
\{aba\}=2(a\circ b)\circ a -a^2\circ b= 2T_aT_b(a)-T_bT_a(a)=(a\circ b)\circ a =0.
\]
 This implies by (i) that $\{s(a)bs(a)\}=0$, and this  in turn implies by Lemma  \ref{le:1.26} that 
  $s(a)\circ b=0$ and by Theorem \ref{th::equiv} that $s(a)\leftrightarrow b$.
\end{proof}

We are now ready to prove the main result of this section.

\begin{theorem}\label{th:jbspect} Let $A$ be a JB-algebra. Then  $A$ is a Rickart if and only if 
$A$ is a spectral.
\end{theorem}

\begin{proof} The family  $(U_p)_{p\in P}$ is a compression base by Corollary \ref{coro:JBcompb}.
If $A$ is Rickart, then it follows by Lemma \ref{le:car} that any $a\in A^+$ has a carrier $s(a)$
(this was proved in \cite[Theorem 2.1, Theorem 2.8]{Arz} for all $a\in A$). By Lemma \ref{lemma:JBproj_cover}, if $a\in
E$, then $s(a)$ is a projection cover of $a$. Hence $(U_p)_{p\in P}$ has the projection cover property.
 We have to prove the comparability property.

For $a\in E$, then by Lemma \ref{lemma:pc}, $s(a)=a^0\in P(a)$. By normalization,  we have $s(a)\in P(a)$ for all $a\in
A^+$.  Let now  $a\in A$, then the elements $a^+, a^-$ are contained in the subalgebra $A(a,1)$, therefore $a^+,a^-\in
C(PC(a))$ by 
\eqref{eq:JB_cpc}. Hence $PC(a)\subseteq PC(\{a^+,a^-\})\subseteq PC(a^+)$ so that 
\[
s(a^+)\in P(a^+)\subseteq C(PC(a^+))\subseteq C(PC(a)). 
\]

Applying Lemma \ref{le:orthog} subsequently to $a^+,a^-$, and then to $s(a^+), a^-$ instead of $a,b$, we obtain first 
that $a^-\in C(s(a^+))$ and $\{s(a^+)a^-s(a^+)\}=0$,
and then that $s(a^+)\leftrightarrow s(a^-)$ and $\{s(a^+)s(a^-)s(a^+)\}=0$.
Since $a^+,a^-\in C(s(a^+))$, we have $a\in C(s(a^+))$.

Put $p:=s(a^+)$, then we proved that  $p\in P(a)$. From $a=a^+-a^-$ we have
\[
U_p(a)=U_p(a^+)-U_p(a^-)=a^+-0=a^+\geq 0,
\]
and
\[
U_{1-p}(a)=U_{1-p}(a^+)-U_{1-p}(a^-)=0-a^-=-a^-\leq 0.
\]
Hence $p\in P^\pm(a)$ and the comparability property is satisfied.

Conversely, assume that $A$ is spectral. Since the sharp elements are precisely the projections, we have by Lemma
\ref{le:principal} and Theorem \ref{th:compressible} that $(U_p)_{p\in P}$ is a spectral compression base.
We will check that the condition (A2) of Definition \ref{de:rickart} is satisfied, so that $A$ is Rickart. Let 
 $a\in A$ and and let $C$ be a C-block of $A$ such that $a\in C$ (Corollary \ref{coro:comparability} (ii)). Then by
\eqref{eq:JB_cpc} we have $A(a,1)\subseteq C$. By Theorem \ref{thm:cblocks} (iv), there is a
 basically disconnected compact Hausdorff space $X$ such that $C\simeq C(X,\mathbb R)$. We will denote the
elements of $C$ and their representing functions in $C(X,\mathbb R)$ by the same symbol. 
Let $Y:=\{x\in X:\ a(x)\ne 0\}$, then the support of $a$ is the closure $\bar Y$, which  is clopen in $X$. 

Let $p$ be the projection in $C$ corresponding to $\bar Y$, then observe that
 $p=a^{**}$. Indeed, we have  $U_p(a)=\tilde U_p(a)= pa=a$, hence $a^{**}\le p$ by Lemma \ref{le:rickmap}. 
Conversely, $a^{**}\in P(a)\subseteq C$, so
that $a=U_{a^{**}}(a)= \tilde U_{a^{**}}(a)=a^{**}a$. Since $\bar Y$ is the support of $a$, this 
  implies that $p\le a^{**}$. Note that it also follows that $a^{**}=(a^2)^{**}$.

 We will  next prove that $^\perp{\{a\}}^+=U_{1-p}(A)^+$, so that (A2) holds.
This  amounts to proving that
\[
\{b\in A^+:\  U_a(b)=0\}=\Ker^+U_p.
\]
Assume first that $b\in \Ker^+U_p$. Since $a=U_p(a)$, we have by \cite[Eq. (1.16)]{AlSh}
\[
U_a(b)=U_{\{pap\}}(b)= U_pU_aU_p(b)=0.
\]
Conversely, let $b\in A^+$ and let $U_a(b)=0$. Then we also have $U_{a^2}(b)=U_aU_a(b)=0$, so that $a^2\perp b$. Let 
\[
c:= a^2-b,
\]
then by the uniqueness of the orthogonal decomposition, we get $a^2=c^+$ and $b=c^{-}$. Consequently, 
if now $C'$ is a block of $A$ containing $c$, then also $a^2,b\in C'$. Considering again the representing functions 
 in $C(X',\mathbb R)$ for a basically disconnected compact Hausdorff space $X'$, we see that
\[
U_{a^2}(b)=0\iff 0=U_b(a^2)=b^2a^2\iff ba^2=0.
\]
It follows that $b$ must be 0 on the support of $a^2$, and we have seen above that the projection corresponding to the
support is $(a^2)^{**}=a^{**}=p$. This implies that $U_p(b)=0$, so that we must have $b\in \Ker^+U_p$.

\end{proof}

\section{Centrally symmetric state spaces}\label{sec:CS}

In this section, we obtain a class of order unit spaces such that $A=V^*$ for a base norm space $V$, where a spectral
compression base exists but the duality of $A$ and $V$ is not spectral.

Let $(X,\|\cdot\|)$ be a normed space and let $B$ be its closed unit ball. 
We will construct a dual pair $(A,V)$ with $A=V^*$  such that the state space $K\cong B$, such state spaces are sometimes called
centrally symmetric (cf. \cite{lami2018ultimate}).

Put $V:= \mathbb R\times X$ and 
\[
V^+:=\{(\alpha,x),\ \|x\|\le \alpha\},\qquad K:=\{(1,x), x\in B\}.
\]
It is easy to see that $V^+$ is a generating  cone: indeed, any element $(\alpha,x)\in V$ can be written as
\[
(\alpha,x)= (\max\{\alpha,\|x\|\},x)-(\max\{\|x\|-\alpha,0\},0)\in V^+-V^+.
\]
Moreover,  $K$ is a base of $V^+$ and is located on the hyperplane 
\[
H_1:=\{(\alpha,x)\in V, \alpha=1\}\not\ni 0.
\]
Note that $co(K\cup -K) =\{(\alpha,x), |\alpha|\le 1, x\in B\}$. This set is clearly radially compact, that is, $B\cap
L$ is a closed segment for every line $L$ through the origin of $V$. By \cite[Prop. II.1.12]{Alf}, $(V,K)$ is a
base normed space and  $co(K\cup -K)$ is the unit ball of the base norm $\|\cdot\|_K$. It follows that
\[
\|(\alpha,x)\|_K=\max\{|\alpha|,\|x\|\}.
\]

Let $A=V^*$, then $A$ is isomorphic to the space $A_b(K)$ of bounded affine functions $K\to \mathbb R$ (\cite{AlSh}).
 With the cone $A^+\cong A_b(K)^+$ of positive functions on $K$ and with $1\cong 1_K$ the constant unit functional,
 $(A,A^+,1)$ is an order unit space, in separating order and norm duality with $V$.

It is easily seen that we may put $A=\mathbb R\times X^*$ with the cone
\[
A^+=\{(a_0,y),\ \|y\|^*\le a_0\}
\]
and the unit $1=(1,0)$. The  order unit norm is
\[
\|(a_0,y)\|_1=\|y\|^*+|a_0|
\]
and the unit interval $E$ has the form
\[
E=\{(a_0,y),\ \|y\|^*\le \min\{a_0,1-a_0\}\},
\]
note that this implies 
\begin{equation}\label{eq:cs_E}
(a_0,y)\in E\ \implies \  a_0\in [0,1]\ \text{ and } \|y\|^*\le 1/2.
\end{equation}

We next characterize some special elements and subsets of $E$ and show their relation to the structure of 
the dual unit ball, which will be denoted by $B^*$.

\begin{lemma}\label{lemma:cs_sharp}
Let $a\in E$,  $a\ne 0$, $a\ne 1$. The following are equivalent.
\begin{enumerate} 
\item[(i)] $a$ is sharp;
\item[(ii)] $\|a\|_1=\|1-a\|_1=1$;
\item[(iii)] $a=(1/2,y)$ with $\|y\|^*=1/2$.
 
\end{enumerate}

\end{lemma}

\begin{proof}
 Assume that $\|a\|_1<1$, then there is some 
 $s\in (0,1)$ such that $(1+s)a\le 1$. But then
\[
sa\le a,\qquad sa\le 1-a,
\]
so that $a$ is not sharp. Since $a$ is sharp iff $1-a$ is sharp, we obtain that (i) implies (ii).

Assume the equality $\|a\|_1=\|1-a\|_1=1$, that is
\[
\|y\|^*+a_0=1=\|-y\|^*+1-a_0
\]
Hence $a_0=1-a_0=1/2=\|y\|^*$ and (ii) implies (iii).

To show that (iii) implies (i), 
let $a=(1/2,y)$ with $\|y\|^*=1/2$ and let $b=(b_0,w)\in E$, $b\le a$. Then $(1/2-b_0,y-w)\in A^+$, so that 
$\|y-w\|^*\le 1/2-b_0$. We obtain
\[
1/2-b_0\le 1/2-\|w\|^*\le |\|y\|^*-\|w\|^*|\le \|y-w\|^*\le 1/2-b_0.
\]
This entails that $b_0=\|w\|^*$ and $1/2-\|w\|^*=\|y-w\|^*$. If also $b\le 1-a$ we similarly obtain
$1/2-\|w\|^*=\|y+w\|^*$. But then 
\[
1/2=\|y\|^*\le 1/2\|y-w\|^*+1/2\|y+w\|^*=1/2-\|w\|^*\le 1/2,
\]
which implies that $b_0=\|w\|^*=0$. Hence $b=0$ and $a$ is sharp.  

\end{proof}

\begin{lemma}\label{lemma:cs_faces}
A subset  $F\subseteq E$ is a face of $E$ not containing 0 or 1 if and only if there is a face $F_0$ of $1/2B^*$ such that
\[
F=\{(1/2,y),\ y\in F_0\}.
\]
In particular, $a\in E$ is an extreme point of $E$ other than 0 or 1 if and only if $a=(1/2,y)$ for some extreme point $y$ of $1/2 B^*$.
\end{lemma}

\begin{proof} Let $F$ be of the given form and let $a=(a_0,y),b=(b_0,w)\in E$ be such that $\lambda a+(1-\lambda) b\in
F$. Then $\lambda a_0+(1-\lambda)b_0=1/2$ and $\lambda y+(1-\lambda)w\in F_0$. Since by \eqref{eq:cs_E} $\|y\|^*,\|w\|^*\le 1/2$, it follows that $y,w\in F_0$ as
well. In particular, $\|y\|^*=\|w\|^*=1/2$, this and \eqref{eq:cs_E} imply that  $a_0,b_0=1/2$. 
This shows that $F$ is a face of $E$, obviously not containing 0 or 1.

Conversely, let $F$ be a face of $E$, $0,1\notin F$,  and let $a=(a_0,y)\in F$. 
Note that if $ta\le 1$ for $t>1$, then $a=t^{-1}ta+(1-t^{-1})0$, so that $0\in F$, similarly, $\|1-a\|_1<1$ implies
$1\in F$, So  we must have $\|a\|_1=\|1-a\|_1=1$ and by Lemma \ref{lemma:cs_sharp}, 
 this implies that $a_0=\|y\|^*=1/2$.
 Let 
\[
F_0:=\{y\in 1/2B^*, (1/2, y)\in F\},
\]
then it is easy to see that $F_0$ is a face of $1/2B^*$. The last statement is now obvious.

\end{proof}

\begin{corollary}\label{coro:cs_onedim} Any extremal element $p\in E$ is one dimensional, that is, $0\le a\le p$ implies 
 $a=tp$ for some $t\in [0,1]$.

\end{corollary}

\begin{proof} By Lemma \ref{lemma:cs_faces} an extremal element in $E$ is of the form 
 $p=(1/2,y)$ for some $y$ extremal in $1/2B^*$.  
Let $0\le b=(b_0,w)\le p$. Since  $\|y\|^*=1/2$, we obtain similarly as
in the proof of Lemma \ref{lemma:cs_sharp} that we must have $\|w\|^*=b_0$ and 
$\|y-w\|^*=1/2-\|w\|^*=\|y\|^*-\|w\|^*$. Put $\lambda:= 2\|y-w\|^*=1-2\|w\|^*$, then we get 
\[
y= \lambda(\frac1{2\|y-w\|^*}(y-w))+ (1-\lambda)(\frac{1}{2\|w\|^*}w).  
\]
As $y$ is extremal, we get $w=2\|w\|^*y=2b_0y$, so that $b=(b_0,2b_0y)=2b_0p$, with $2b_0\in [0,1]$.

\end{proof}

We next describe retractions, F-compressions and compressions on $A$. For this, we need to introduce the following
notations for $x\in X$ and $y\in X^*$:
\[
\partial_x:=\{y\in B^*,\ \langle y,x\rangle=\|x\|\},\qquad \partial^*_y:=\{x\in B, \ \langle y,x\rangle=\|y\|^*\}.
\]
Then $\partial_x$, $\partial^*_y$ are faces of the respective unit ball and  $\partial_x\neq \emptyset$.

\begin{prop}\label{prop:cs_retractions}
 An element $p\in E$ is the focus of a retraction if and only if 
$p=(1/2,y)$, where $y$ is an extremal element in $1/2B^*$ attaining its norm on $B$ (i.e. $\partial^*_y\ne \emptyset$). 
 A map $J:A\to A$ is a retraction with focus $p$ if an only if 
\[
J(a)=(a_0+\langle x,w\rangle)p,\qquad a=(a_0,w)\in A
\]
for some $x\in \partial^*_y$. Moreover
\begin{enumerate}
\item[(i)] The map $J$ is an F-compression if and only if $\partial_x$ is a singleton, that is, $\partial_x=\{2y\}$.
\item[(ii)] The map $J$ is a compression if and only if $\partial^*_y=\{x\}$ and $\partial_x=\{2y\}$.
\end{enumerate}
\end{prop}

\begin{proof}
Assume that $J:A\to A$ is a retraction with focus $p$. By Lemma \ref{lemma:princ_ext_sharp}, $p$ must
be extremal, so that $p=(1/2,y)$ for an extremal element $y\in 1/2B^*$. Note that $1-p= (1/2,-y)$, so that,
by symmetry of $B^*$, $1-p$ is an extremal element in $E$ as well. It follows that both $p$ and $1-p$ are one
dimensional. By \eqref{eq:imj}, we obtain 
 $\Iim^+J= \mathbb R^+p$ and since $\Iim(J)$ is positively generated, $\Iim(J)=\mathbb R p$.
Consequently, $J$ must be of the form
\[
J(a)=\langle a,\rho\rangle p,\qquad a\in A,
\]
for some linear functional $\rho$ on $A$. Since $J$ is continuous, positive and $J(1)=p$, we must have $\rho\in K$ and 
 since $J$ is idempotent, $\langle p,\rho\rangle =1$, so that 
\[
\rho\in K_p=\{(1,x),\ x\in B,\ \langle (1/2,y), (1,x)\rangle=1\}=\{(1,x),\ x\in \partial^*_y\}. 
\]
Let $0\le a\le p$, then by Corollary \ref{coro:cs_onedim}, $a=tp$ for some $t\in [0,1]$ so that
\[
J(a)=tJ(p)=tp=a.
\]
Hence a map of this form is indeed a retraction. This proves the first part of the statement.

A retraction is an F-compression if and only if it satisfies (F3) of Definition \ref{de:Fcompr}. So let $a=(a_0,w)\in E$,
 then clearly $J(a)=0$ if and only if $a_0+\langle x,w\rangle=0$, which means that $-a_0^{-1}w\in \partial_x$. So 
(F3) is true if and only if such $a$ is in $[0,1-p]$, but since $1-p$ is one-dimensional, this means that
$(a_0,w)=t(1-p)=(t/2,-ty)$, which implies  $-a_0^{-1}w=2y$. Since 
we always have $2y\in \partial_x$, this proves (i).

By the symmetry of the unit ball, we see from (i) that if $J$ is an F-compression, then it has a complementary
F-compression $J'$ with focus $1-p$. Indeed, if $x\in \partial^*_y$ and $\partial_x=\{2y\}$, then clearly $-x\in
\partial^*_{-y}$ and $\partial_{-x}=\{-2y\}$. By Theorem \ref{thm:AS_FP}, $J$ is a compression if and only if both $J$
and $J'$ are smooth. 
By Lemma \ref{lemma:smooth}, $J$ is smooth if and only if
\[
K_p=\mathcal S(J)=\Iim(J^*)\cap K.
\] 
Since $J^*\varphi=\langle p,\varphi\rangle \rho$, we see that 
\[
\mathcal S(J)=\{\omega\in K:\ J^*\omega=\omega\}=\{\omega\in K:\ \langle p,\omega\rangle \rho=\omega\}=\{\rho\}.
\]
This means that $K_p$ and consequently also  $\partial^*_y$ must be a  singleton, and  $\partial^*_y=\{x\}$. If this is
true, $J$ is smooth and again by symmetry, $J'$ is smooth as well. This proves
(ii).

\end{proof}

A normed space $X$ is called strictly convex if its closed unit ball $B$  is strictly convex, that is, 
every boundary point of $B$ is an extreme point of $B$. A normed space is called smooth if any nonzero element 
attains its norm at a unique point of the dual unit ball. 
If $X^*$ is strictly convex (smooth), then $X$ is smooth (strictly convex). If $X$ is reflexive, then these properties
are mutually dual. More precisely, 
$X$ is smooth iff $X^*$ is strictly convex  iff $\partial_x$ is a singleton for all $0\le x\in X$.  
For more details, see e.g. \cite[\S 26]{kothe}.

\begin{theorem}\label{thm:cs_spectral} Let $(X,\|\cdot\|)$ be a Banach space and let the pair $(A,V)$ be constructed as
above. Then there is a spectral compression base $(J_p)_{p\in P}$ on $A$ if and only if $X$ is  reflexive and smooth.

\end{theorem}

\begin{proof} Let $(J_p)_{p\in P}$ be a spectral compression base. By Lemma \ref{le:principal}, all sharp elements are
in $P$. By Lemma \ref{lemma:cs_sharp} and Proposition \ref{prop:cs_retractions}, this means that
$\partial^*_y\ne\emptyset$ for all $y\in X^*$ with $\|y\|^*=1/2$, and hence for all nonzero $y$ in $X^*$. By James's
theorem, this means that $X$ is reflexive. Moreover, since for any $y$ with $\|y\|^*=1/2$, 
 $\{2y\}=\partial_x$ is a face of $B^*$, we see that $X^*$ must be strictly convex, so that $X$ is smooth.

Conversely, if $X$ is a reflexive smooth Banach space, then $\partial^*_y\ne \emptyset$ for all $y\ne 0$ and 
 $\partial_x$ is a singleton for all $x\ne 0$. Therefore,  any $p=(1/2,y)$ with $\|y\|^*=1/2$ is a focus of an
F-compression $J_p$ of the form 
\[
J_p(a)=(a_0+\langle x_p,w\rangle)p,\qquad a=(a_0,w),
\]
for some choice of $x_p\in \partial^*_y$. 

Let $P=\{(1/2,y),\ \|y\|^*=1/2\}\cup\{0,1\}$ be the set of all sharp elements in $E$. Since $B^*$ is strictly convex, 
 all $p\in P$, $p\ne 0,1$  are extremal, hence one dimensional. Using this fact, it is easily seen that $P$ is a normal subalgebra in
$E$ and that $(J_p)_{p\in P}$ is a compression base with the projection cover property (we put $J_1=id$ and $J_0=0$). 
To prove comparability, let $a=(a_0,w)\in A$. If $\pm a\in  A^+$, then clearly $0$ or $1$ is in $P^\pm(a)$. So assume that
$\pm a\not\in A^+$, which amounts to $\|w\|^*>|a_0|$. Put $y:=(2\|w\|^*)^{-1}w$, then $p=(1/2,y)\in P$ and we have
\[
(a_0,w)=(\|w\|^*+a_0)p-(\|w\|^*-a_0)(1-p),
\]
hence $p\in P^\pm(a)$. So $(J_p)_{p\in P}$ is a spectral compression base.

\end{proof}

\begin{theorem}\label{thm:cs_spectralduality} Let $(X,\|\cdot\|)$ be a Banach space and let the pair $(A,V)$ be
constructed as above. Then $(A,V)$ is in spectral duality if and only if $X$ is reflexive, smooth and strictly convex.

\end{theorem}

\begin{proof} Assume that $(A,V)$ are is spectral duality. Let $\mathcal P$ be the set of all projective units and let
$J_p$ be the corresponding compressions. By Theorem \ref{th:gencomp}, $(J_p)_{p\in \mathcal P}$ is a spectral
compression base, so that by Theorem \ref{thm:cs_spectral}, $X$ is reflexive and smooth. As before, all sharp elements
are in $\mathcal P$, so that $(1/2,y)$ with $\|y\|^*=1/2$ is a projective unit. By Proposition \ref{prop:cs_retractions}, 
$\partial^*_y$ must be a singleton, which implies that $X^*$ is smooth. Hence $X$ is strictly convex.

Conversely, let $X$ be a strictly convex and smooth reflexive Banach space. Then we may construct a spectral compression
base $(J_p)_{p\in P}$ as in the proof of Theorem \ref{thm:cs_spectral}, where $P$ is the set of all sharp elements. 
Since $\partial^*_y$ is a singleton for all $y\ne 0$, it follows by Proposition \ref{prop:cs_retractions} that all $J_p$
are compressions. The proof now follows by Theorem \ref{th:gencomp}.

\end{proof}

\section{Conclusions}

We studied in some detail the notion of  spectrality in order unit spaces in the sense of Foulis \cite{FPspectres}.
This notion is based on a particular set of idempotent mappings, called a compression base. Spectrality is characterized
by two properties of the compression base: projection cover property and comparability. While the projection cover
property can be trivial in some cases (consider for example the compression base consisting only of the zero and
identity maps), we proved that the comparability property is already rather strong. If comparability holds, the order unit space
is covered by subsets called C-blocks that admit a functional representation. Moreover, this property is equivalent to
spectrality if the set $P$ of projections is monotone $\sigma$-complete, in particular, in the finite dimensional case.

We compared this notion of spectrality with the spectral theory of Alfsen and Shultz \cite{AlSh}, where it is assumed 
that the order unit space is the dual of a Banach space. We proved that this theory is a
special case of the Foulis spectrality and found conditions under which they coincide, as well as examples when they are
different. We studied the spectral theory for JB-algebras and proved that Rickart JB-algebras are exactly those that are 
spectral in the Foulis sense.  In this case, the compression base is formed by compressions in the sense of Alfsen-Shultz, but the Alfsen-Shultz theory is not applicable since a Rickart JB-algebra does not have a predual in general. 

There is also a more general 
 version of Alfsen-Shultz theory \cite{AS}, where $A$ is only assumed to be monotone $\sigma$-complete. For C*-algebras,
it is known that a C*-algebra   is Rickart if and only if the self-adjoint part of any maximal abelian subalgebra is
monotone $\sigma$-complete \cite{wright}, but it is not known if this implies monotone $\sigma$-completeness for the
self-adjoint part of the whole algebra. A similar characterization of Rickart JB-algebras is unknown. Here the maximal
abelian subalgebras should be replaced by maximal associative subalgebras. 
It is an interesting question whether the maximal associative subalgebras of Rickart JB-algebras coincide with the
C-blocks.

\end{document}